\documentclass[nonacm,sigconf,screen, dvipsnames]{acmart}
\usepackage{graphicx}
\usepackage{dcolumn}
\usepackage{bm}
\usepackage{hyperref}
\usepackage{tikz}
\usepackage{pgfplots}
\usepackage{placeins}
\usepackage{tikz-3dplot}
\usepackage{float}
\usepackage{subcaption}
\usepackage{adjustbox}
\usepackage{quantikz}
\usetikzlibrary{shapes.geometric, arrows}

\tikzstyle{prep_sense} = [rectangle, rounded corners, minimum width=2.5cm, minimum height=0.8cm, text centered, draw=black, fill=blue!20]
\tikzstyle{storage_retrieval} = [rectangle, rounded corners, minimum width=2.5cm, minimum height=0.8cm, text centered, draw=black, fill=green!20]
\tikzstyle{memory_style} = [rectangle, rounded corners, minimum width=2.5cm, minimum height=0.8cm, text centered, draw=black, fill=red!20]
\tikzstyle{processing} = [rectangle, rounded corners, minimum width=2.5cm, minimum height=0.8cm, text centered, draw=black, fill=violet!20]
\tikzstyle{arrow} = [thick,->,>=stealth]

\AtBeginDocument{%
  }

\begin{document}

\title{STQS: A Unified System Architecture for Spatial Temporal Quantum Sensing}

\author{Anastashia Jebraeilli}
\affiliation{
    \institution{Pacific Northwest National Lab}
    \city{Richland}
    \state{Washington}
    \country{USA}
}
\affiliation{
    \institution{University of Georgia}
    \city{Athens}
    \state{Georgia}
    \country{USA}
}
\email{anaj@uga.edu}
\orcid{0000-0002-8725-2088}

\author{Chenxu Liu}
\affiliation{
  \institution{Pacific Northwest National Lab}
  \city{Richland}
  \state{Washington}
  \country{USA}
}
\email{chenxu.liu@pnnl.gov}
\orcid{0000-0003-2616-3126}

\author{Keyi Yin}
\affiliation{
  \institution{University of California, San Diego}
  \city{San Diego}
  \state{California}
  \country{USA}
}
\email{keyin@ucsd.edu}
\orcid{0009-0005-7563-271X}

\author{Erik W Lentz}
\affiliation{
  \institution{Pacific Northwest National Lab}
  \city{Richland}
  \state{Washington}
  \country{USA}
}
\email{erik.lentz@pnnl.gov}
\orcid{0000-0001-7807-4825}

\author{Yufei Ding}
\affiliation{
  \institution{University of California, San Diego}
  \city{San Diego}
  \state{California}
  \country{USA}
}
\email{yufeiding@ucsd.edu}
\orcid{0000-0002-8716-5793}

\author{Ang Li}
\affiliation{
  \institution{Pacific Northwest National Lab}
  \city{Richland}
  \state{Washington}
  \country{USA}
}
\affiliation{
  \institution{University of Washington}
  \city{Seattle}
  \state{Washington}
  \country{USA}
}
\email{ang.li@pnnl.gov}
\orcid{0000-0003-3734-9137}

\begin{abstract}

Quantum sensing (QS) harnesses quantum phenomena to measure physical observables with extraordinary precision, sensitivity, and resolution. QS is among the most promising avenues for quantum technologies, alongside quantum computing and networking. Despite significant advancements in quantum sensing, prevailing efforts have focused predominantly on refining the underlying sensor materials and hardware. Given the growing demands of increasingly complex application domains and the continued evolution of quantum sensing technologies, the present moment is the right time to systematically explore distributed quantum sensing architectures and their corresponding design space.     

We present STQS, a unified system architecture for spatiotemporal quantum sensing that interlaces four key quantum components: \emph{sensing}, \emph{memory}, \emph{communication}, and \emph{computation}. 
By employing a comprehensive gate-based framework, we systemically explore the design space of quantum sensing schemes and probe the influence of noise at each state in a sensing workflow through simulation. We introduce a novel distance-based metric that compares reference states to sensing states and assigns a confidence level. We anticipate that the distance measure will serve as an intermediate step towards more advanced quantum signal processing techniques like quantum machine learning. To our knowledge, STQS is the first system-level framework to integrate quantum sensing within a coherent, unified architectural paradigm. STQS provides seamless avenues for unique state preparation, multi-user sensing requests, and addressing practical implementations. We demonstrate the versatility of STQS through evaluations of quantum radar and qubit-based dark matter detection. To highlight the near-term feasibility of our approach, we present results obtained from IBM’s Marrakesh and IonQ's Forte devices, validating key STQS components on present day quantum hardware.

\end{abstract}

\maketitle

\section{Introduction}\label{sec:intro}

Recent advances in quantum sensing have markedly advanced the precision of measurement to address the growing need for new sensors capable of richer acquisition with reduced power consumption \cite{passian2022concept}. In particular, the intrinsic hypersensitivity of Noisy Intermediate-Scale Quantum (NISQ) era qubits \cite{Preskill_2018} to subtle environmental perturbations has proven to be a promising quantum sensor platform. By exploiting the delicate phenomena of quantum coherence and decoherence, qubit-based sensors can achieve precision and robustness surpassing even the most fundamentally idealized classical sensors \cite{Zhang_2021}. Such advances promise to overcome the long-standing limitations in sensor design \cite{doi:10.1126/sciadv.ade4454, kuffer2024sensingoutofequilibriumquantumnongaussian}.

Despite significant progress, the multipartite architecture of quantum sensing systems and inherently burdensome need for manual, experiment-specific adjustments continue to impede systematic design and optimization efforts. Due to the lack of a comprehensive framework to coherently integrate sensor operation, memory, communication, and computation, the scalability and precision of quantum sensors have been largely limited. Current approaches frequently fail to capture how noise accumulates and propagates through a sensing protocol, complicating theoretical modeling and hindering real-world utilization of quantum sensors.

These challenges need a thorough reexamination from a computer architecture viewpoint to address how the elements of a distributed quantum sensing framework interact and depend on one another. In this spirit, we assert that an effective system-level design for quantum sensing should leverage classical architectural principles, like memory hierarchies, data flow optimizations, and noise-robust communication protocols, while also incorporating quantum-specific mechanisms and techniques for entanglement distribution, state preparation, and error correction. 

We introduce a unified system architecture for Spatial Temporal Quantum Sensing (STQS). As illustrated in Figure \ref{fig:QSS}, STQS provides a comprehensive framework for distributed quantum sensing, seamlessly integrating noise modeling and error mitigation at every stage of the sensing protocol. In contrast to existing approaches where infrastructure, architectural design, and optimization methods are primarily treated as separate concerns, STQS establishes a new paradigm for automatically determining task-specific quantum sensing configurations. Rather than aggregating existing tools, STQS obviates labor-intensive manual calibration and fine-tuning to enable systemic exploration of the expansive design space underling the next generation of quantum sensors.

STQS accommodates a broad class of quantum infrastructures by translating diverse sensing architectures into a discrete-variable (DV) simulational framework. Using STQS, simulations of continuous-variable (CV) sensing processes can be modeled as DV processes (enabling the application of existing DV noise models). The generality of STQS not only removes a bottleneck for researchers who previously had to painstakingly tailor infrastructures to accommodate automated optimization but also incorporates a robust representation of the noise and errors inherent to a sensing modality. Consequently, STQS effectively streamlines the design and refinement process, enabling simulations that guide researchers toward more efficient hardware networking. 

Figure \ref{fig:stqs-circuit} illustrates the STQS pipeline. Formulated as a unified gate-based framework for investigating distributed quantum sensing protocols, STQS abstracts sensing operations as standard quantum gates, such as phase, rotation, swap, and CNOTs. STQS enables systematic exploration of a sensing scheme's performance under diverse experimental settings. STQS's seamless integration of quantum sensing with quantum computing supports optimizing sensing protocols with simulations that accurately capture the interplay of noise, quantum state preparation, post-processing, resource constraints, and error correction. Prior system-level efforts in quantum information science have primarily targeted quantum computing architectures and networks, leaving a notable gap in the literature regarding distributed quantum sensing protocols \cite{doi:10.1126/science.aam9288, MURALI2019102, Cacciapuoti_2024, LaRose_2019}. To our knowledge, this work constitutes the first system-level study of distributed quantum sensing protocols, offering a bridge between infrastructure-agnostic simulations and practical, noise-aware experimental implementations. 

We begin by providing a concise overview of the fundamentals of quantum sensing and their relevant noise models in Section \ref{sec:background}. In Section \ref{sec:unified-framework}, we present the specifics of the STQS architecture and pipeline, introducing prediction metrics grounded in similarity measurements that we believe have high relevance to quantum machine learning. To further demonstrate the adaptability and versatility of STQS, in Section \ref{sec:applications}, we present two concrete applications: quantum radar and dark matter detection, followed by results from quantum hardware in Section \ref{sec:hardware}, validating the key design principles of STQS—however, STQS remains a proposal for the simulation-based study. By unifying the sensing, memory, and computing layers within a gate-based framework, we can rigorously explore the interplay of noise, entanglement distribution, and resource constraints.

In summary, we will introduce a quantum sensing framework that unifies spatiotemporal considerations within distributed quantum sensing protocols. By tackling the complexity of optimizing the design space for quantum sensing systems, STQS efficiently mitigates the computational and experimental burdens that have, until now, hindered progress. We assert that our work makes the following contributions:
\begin{itemize}
    \item \textbf{Unified STQS architecture: } A novel integrated system for spatially and temporally distributed quantum sensing, combining sensing, memory, computation, post-processing, and communication within a single cohesive framework.
    \item \textbf{Gate-based modeling and optimization: } A gate-based formalism to characterize and optimize distributed quantum sensing protocols in the presence of realistic noise sources.
    \item \textbf{Effective approach for sensed quantum signal classification:} A simple and practical approach to classify sensed quantum states with respect to a specific but possibly unknown reference state prepared by users, where the confidence level can be quantized, ensuring accurate classification and security of the user-provided state.
    \item \textbf{Design insights for future experiments:} We offer guidance on the interplay of noise, resource allocation, and sensing efficiency to provide useful hints for the next generation of quantum sensing research. 
\end{itemize}

\section{Background}\label{sec:background}

\subsection{Quantum Sensing}\label{sec:quantum sensing background}

Unlike classical sensing, quantum sensing uses quantum resources to measure physical quantities with precision beyond the capabilities of classical instruments. 
With recent advancements in quantum technology, several physical platforms are now being used as quantum sensors. Nitrogen-vacancy (NV) centers in diamonds are commonly employed for magnetometry \cite{Taylor_2008}, while other popular quantum sensing platforms include Rydberg atoms \cite{Tu2023ApproachingTS}, superconducting qubits \cite{Devoret2013SuperconductingCF}, and trapped ions \cite{Schmidt:2005tpj}. Additionally, photons prepared in entangled states, such as NooN or squeezed states, are also widely used for various quantum sensing tasks \cite{Lee_2016}. However, the parameter space for designing an individual sensor and its corresponding distributed sensing protocol is vast, highly detailed, and variable. Each sensing platform, configuration, and measurement task presents a unique set of parameters that users must finely-tune to ensure maximum information gain while maintaining the integrity of the quantum resources. A typically iterative design process further amplifies this non-trivial complexity. Consequently, there exists an increasing need within the quantum sensing community to make emerging technologies more accessible and scalable.

STQS addresses this need by addressing the significant gap in the architectural landscape of quantum sensing. STQS simplifies the design and optimization process, enabling comprehensive evaluations of how best to enhance a sensing protocol's performance under various conditions.

\subsection{Standard Quantum Limit and Heisenberg Limit }\label{sec:HL vs SQL}

One of a sensing protocol's key metrics is the accuracy and precision of the measurement results. It is well understood that using more sensors to measure the same physical quantity can enhance sensing precision. However, for classical sensing that does independent measurements, statistical theory predicts that the improvement in precision is limited by a factor of $\sqrt{N}$ where $N$ is the number of sensors (or samplings). This standard quantum limit (SQL) represents the best achievable precision in classical sensing protocols when excluding all other experimental imperfections.

However, in quantum sensing, when the sensors are prepared in an entangled state, the detection scheme's measurement precision can surpass the SQL. Maximizing the quantum Fisher information~\cite{PhysRevLett.102.100401}, can improve the precision by a factor of $N$, known as the Heisenberg limit (HL). This represents the ultimate precision achievable with quantum sensing, offering a significant advantage over classical methods. 

Here, we derive the error propagation for estimating the parameter of interest under idealized, noiseless conditions, considering both unentangled and entangled sensing probes. This derivation recovers the SQL for independent measurements and establishes the HL when entangled probes are used, demonstrating the inherent quantum advantage over classical sensing methods.

Let us begin by addressing the unentangled probes $N$, which we will show are limited in uncertainty by a scaling of $\frac{1}{\sqrt{N}}$. An unentangled probe is initially in the state,
\begin{equation}
|\psi_{\text{initial}}\rangle = \frac{1}{\sqrt{2}}(|0\rangle + |1\rangle).
\end{equation}
When an external parameter $\phi$ is accrued onto the sensing subspace, it acts on the state of the sensor to transform it to:
\begin{equation}
|\psi_\text{final}\rangle = \frac{1}{\sqrt{2}}(|0\rangle + e^{i\phi}|1\rangle).
\end{equation}
The probability $p(\phi)$ that $|\psi_{\text{initial}}\rangle = |\psi_\text{final}\rangle $ is
\begin{equation}
p(\phi) = |\langle\psi_\text{initial}|\psi_\text{final}\rangle|^2 = \cos^2(\frac{\phi}{2}).
\end{equation}
So, we can compute the sensed parameter from the probability by:
\begin{equation}
\phi = 2\arccos(\sqrt{p(\phi)}),
\end{equation}
with $N$ independent measurements, each with $p(\phi)$ of successful outcomes, there is a binomial distribution
\begin{equation}
P(k | N, p(\phi)) = \binom{N}{k} (p(\phi))^k (1 - p(\phi))^{N - k},
\end{equation}
where $k$ is the number of successful measurements (i.e., $|\psi_{\text{initial}}\rangle = |\psi_\text{final}\rangle $), with a mean $\langle k \rangle = Np(\phi)$, and the variance in successes is $\Delta^2 k = Np(\phi)(1-p(\phi))$. Also note that $p(\phi) \approx \frac{k}{N}$.
So, we write the uncertainty in probability $p(\phi)$ due to finite samplings N as
\begin{equation}
\Delta p(\phi) = \frac{\sqrt{\Delta^2k}}{N} = \frac{\sqrt{N(\phi)(1-p(\phi))}}{N},
\end{equation}
\begin{equation}
\Delta p(\phi) = \sqrt{\frac{p(\phi)(1-p(\phi))}{N}} = \sqrt{\frac{\cos^2(\frac{\phi}{2}(1-\cos^2(\frac{\phi}{2})))}{N}}.
\end{equation}
The error propagated onto the computation of $\phi$ is $\Delta\phi$,
\begin{equation}
\Delta(\phi) = \frac{\Delta p(\phi)}{|\frac{\partial p(\phi)}{\partial \phi}|},
\end{equation}
which simplifies to $\Delta \phi = \frac{1}{\sqrt{N}}$ because,
\begin{equation}
|\frac{\partial p(\phi)}{\partial \phi}| = 
|\frac{\partial}{\partial \phi} * \cos^2(\frac{\phi}{2})| = |\sin(\frac{\phi}{2})*(-\cos(\frac{\phi}{2}))|, 
\end{equation}
which yields,
\begin{equation}
\Delta(\phi) = \frac{\Delta p(\phi)}{|\frac{\partial p(\phi)}{\partial \phi}|} = \frac{\sqrt{\frac{\cos^2(\frac{\phi}{2})(1-\cos^2(\frac{\phi}{2}))}{N}}}{|\sin(\frac{\phi}{2})*(-\cos(\frac{\phi}{2}))|} = \sqrt{\frac{1}{N}}.
\end{equation}
This is, by definition, the standard quantum limit. The uncertainty in calculating $\phi$ from the probability distribution of $p(\phi)$ scales only with the number of independent measurements.

Let's see how sensing with entangled probes is bound by the Heisenberg limit, $\frac{1}{N}$ scaling. The entangled probe GHZ-state spanning from sensing qubits $q_1$ to $q_N$ is initially in the state 
\begin{equation}
|\psi_{\text{initial}}\rangle = \frac{1}{\sqrt{2}}(|0\rangle^{\otimes N} + |1\rangle^{\otimes N}).
\end{equation}
After sensing the unknown parameter $\phi$, we have  
\begin{equation}
|\psi_\text{final}\rangle = \frac{1}{\sqrt{2}}(|0\rangle^{\otimes N} + e^{i N\phi }|1\rangle^{\otimes N}),
\end{equation}
with a probability $p(\phi)$ that $|\psi_{\text{initial}}\rangle = |\psi_\text{final}\rangle $ of
\begin{equation}
p(\phi) = |\langle\psi_\text{initial}|\psi_\text{final}\rangle|^2 = cos^2(\frac{N\phi}{2}),
\end{equation}
where $N$ is the number of probes used, with an estimated error of 
\begin{equation}
\Delta^2p(\phi) = \langle\psi_\text{final}|(|\psi_\text{initial} \rangle \langle \psi_\text{initial}|)^2 |\psi_\text{final}\rangle - p^2(\phi),
\end{equation}
leading to an error in $\phi$ of, 
\begin{equation}
\Delta(\phi) = \frac{\Delta p(\phi)}{|\frac{\partial p(\phi)}{\partial \phi}|} = \frac{1}{N},
\end{equation}
demonstrating a clear advantage over the unentangled case by a factor of $\frac{1}{\sqrt{N}}$.

\subsection{Modeling Noise for Quantum  Sensing}\label{sec:noise model}

Various factors, including the design of the sensing protocol, instrumentation, imperfections, and noise, influence the precision and accuracy of quantum sensing measurements. In this work, we model the quantum sensing processes by using quantum circuits and by incorporating noise channels into the circuits to simulate the types of noise that can occur in sensing applications. The noise models we consider include (1) depolarization, (2) thermal relaxation, (3) dephasing, and (4) measurement errors.

Depolarization errors are described by the quantum channel,
\begin{equation}
    \mathcal{E}_\text{dep}(\rho) = (1-p) \rho + p \ I/2^n,
\end{equation}
where $n$ is the number of qubits. Thermal relaxation errors are modeled by the quantum channel,
\begin{equation}
    \mathcal{E}_\text{th}(\rho) = \left( 
    \begin{array}{cc}
       \rho_{00} + (1 - e^{-t/T_1}) \rho_{11}  & e^{-t/(2T_2)} \rho_{01} \\
       e^{-t/(2T_2)} \rho_{10}  & e^{-t/T_1} \rho_{11}
    \end{array}
    \right),
\end{equation}
where $T_1$ and $T_2$ are the qubit relaxation and decoherence times, and $t$ is the quantum operation's duration. Dephasing is represented by
\begin{equation}
    \mathcal{E}_{z}(\rho) = (1-p) \rho + p\ Z \rho Z,
\end{equation}
where $Z$ is a Pauli Z operation. Finally, measurement (readout) errors occur when a qubit in $|0\rangle$ is measured as $|1\rangle$ ($p_{0\rightarrow1}$), or when a qubit in $|1\rangle$ is measured as $|0\rangle$ ($p_{1\rightarrow0}$).

STQS's noise modeling method maps CV-based sensing setups to a DV-based pipeline. Although previous work has described mapping noise in CV systems to DV systems \cite{PhysRevLett.102.120501, PhysRevA.64.012310, PhysRevX.6.031006}, our contribution is the first to fully model these noise dynamics within a unified framework for spatiotemporal quantum sensing. The validity of this mapping is supported by the well-established equivalence of noise evolution in CV and DV systems under the Lindblad master equation framework. The Lindbladian governs the evolution of open quantum systems and provides a unified description of decoherence and dissipation independently of whether the underlying Hilbert space is discrete or continuous. Since quantum noise manifests as a dynamical process that can be modeled as a quantum channel, the mathematical form of the Lindbladian remains unchanged, whether describing phase-space diffusion in a CV system or depolarizing noise in a DV system \cite{preskill1998lecture}. The fundamental structure of quantum noise, dictated by entirely positive trace-preserving (CPTP) maps, ensures that noise processes in CV systems can always be systematically translated into an equivalent DV representation \cite{Gregoratti01042003}. Furthermore, many foundational quantum error correction works have established rigorous theoretical frameworks for stabilizing quantum information in CV systems and DV encodings. The Gottesman-Kitaev-Preskill (GKP) encoding \cite{PhysRevA.64.012310} in particular, provides a compelling demonstration of how errors in CV systems can be effectively mapped onto a DV error model to enable fault-tolerant quantum error correction within bosonic systems. This mapping converts small continuous displacements in phase space into discrete Pauli errors on an encoded logical qubit. Essentially, \cite{PhysRevA.64.012310} shows how, by discretizing CV noise into Pauli $X$ and $Z$ errors, there is a direct connection between noise in CV and DV systems. We are simply the first to apply the well-documented and studied mapping of noise in CV systems to noise channels in DV systems for distributed quantum sensing. We implore the reader who may still be skeptical to look at \cite{preskill1998lecture, Gregoratti01042003, PhysRevLett.102.120501, PhysRevA.64.012310, PhysRevX.6.031006} and related works to convince themselves of the validity of this mapping.

The properties of the physical platforms we included in this manuscript are summarized in Table~\ref{tab:qubit_errors2}, and this information is referred to as the default noise profile in the simulations presented in this work. We note that the results of Table \ref{tab:qubit_errors2} are a compilation of current error rates for various quantum system platforms, and are a valuable resource for the sensing community, as a consolidation of critical performance metrics into an easily accessible and digestible format.

\begin{table}
    \centering
    \caption{Performance characteristics of NISQ qubits used for building the noise model of STQS. SGE: single-qubit gate error rate, TGE: two-qubit gate error rate, SPE: state preparation error rate. ME: measurement error rate. }
    \centering
    \scalebox{0.9}{
    \begin{tabular}{c|c|c|c|c}
    \toprule

         &   Trapped ions & Rydberg atoms &  SC Qubits & NV centers\\
         \midrule
         T1&   min-hr &10-100s $\mu$s  &  10-200 $\mu$s & 1-10 ms \\
         
         T2&   ms-s  &10-100 $\mu$s  &  10-300 $\mu$s & 10-100 $\mu$s \\
         
         SGE &   <1\%  & 0.1-1\% &  <0.1\%  & <1\% \\
         
         TGE &   1-2\% &1-5\% &  1-2\% & 1-5\%\\
         SPE &   <1\% &1-5\% &  1\% & <1\% \\
         ME &   1-2\% &1-10\%  &  1-5\% & 1-10\%\\
        
         Readout &   30-100 $\mu$s &1-10 $\mu$s &  100-500 ns & 1-10 $\mu$s\\
         \bottomrule
    \end{tabular}
    }
    \label{tab:qubit_errors2}
\end{table}

\section{STQS Framework}\label{sec:unified-framework}

In the subsequent subsections, we detail the system architecture and pipeline that form the backbone of the STQS framework. This description encompasses the modular design and sequential operational stages, from probe preparation and signal acquisition to data storage and post-processing. These elements help further explore the framework’s capabilities and adaptability to a wide array of sensing applications.
\subsection{System Architecture and Pipeline}\label{sec:overall}

\begin{figure}[h]
        \centering
    \includegraphics[width=0.95 \linewidth]{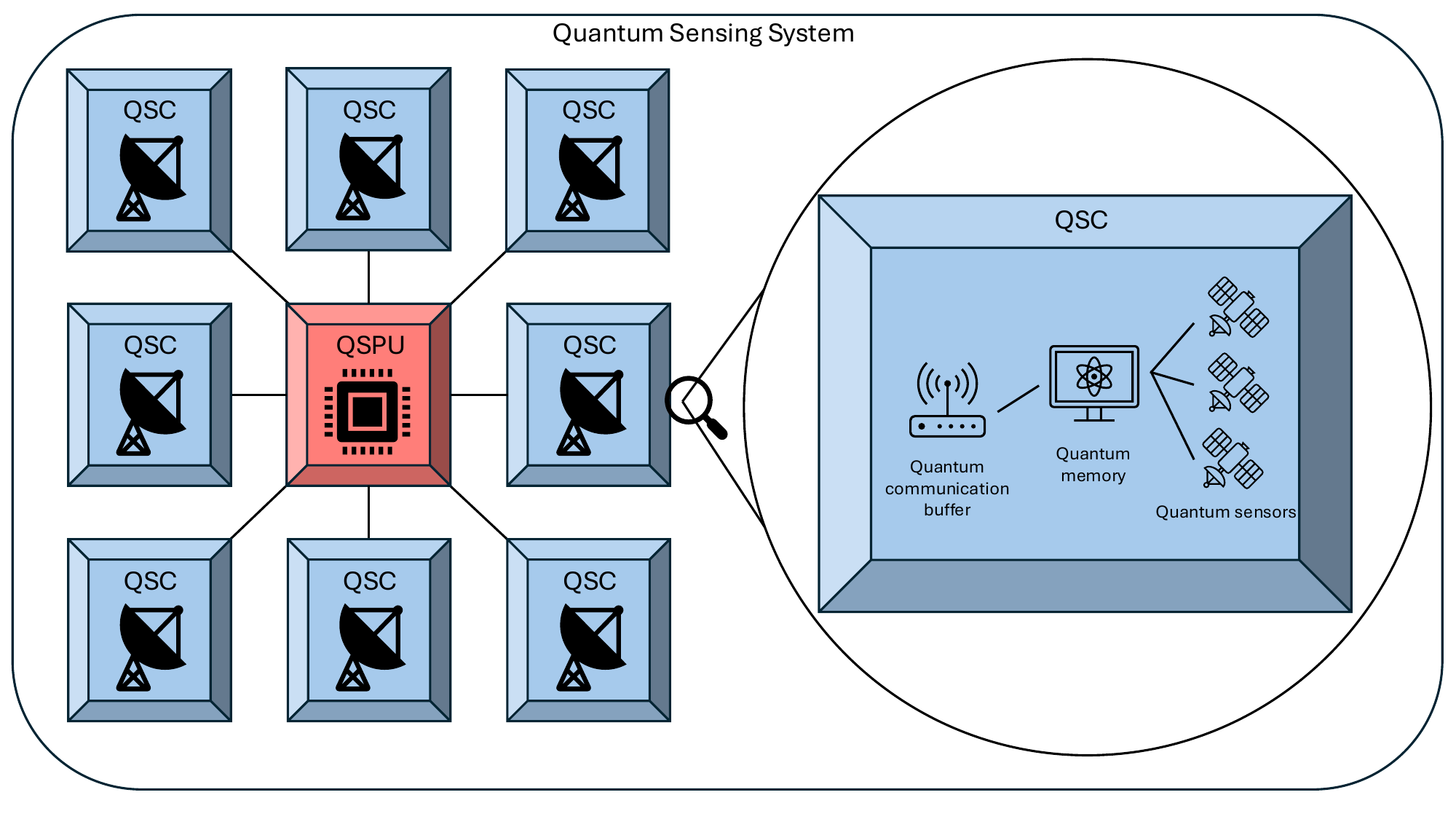}
        \caption{ System architecture for distributed quantum sensing using STQS. The inset diagram is an enlargement of a single QSC to depict the components within the network more clearly.} 

        \label{fig:QSS}
\end{figure}
    
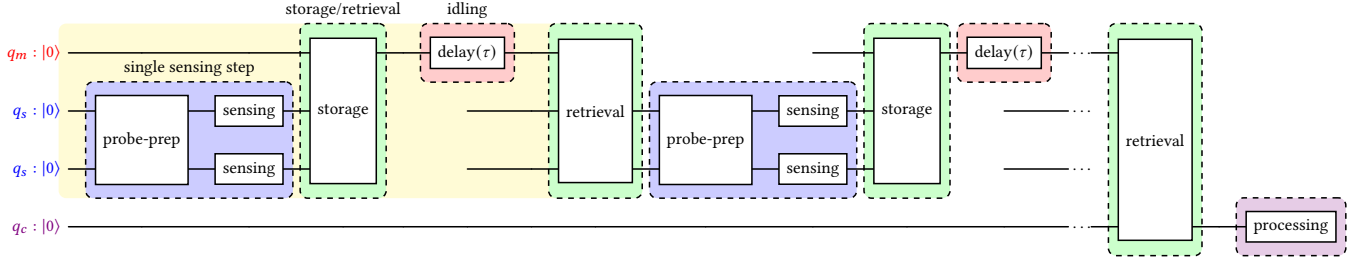
\begin{figure*}[h]
        \centering
        \resizebox{\linewidth}{!}{%
    \begin{quantikz}
        \lstick[label style={red}]{$q_m: {|0\rangle}$}
        \gategroup[3,steps=8,style={draw=none, rounded corners,fill=yellow!20, inner xsep=2pt},background,label style={label position=above,anchor=north, yshift=0.2cm}]{}
        & & &\gate[3]{\text{storage}}\gategroup[3,steps=1,style={dashed,rounded corners,fill=green!20, inner xsep=2pt},background,label style={label position=above,anchor=north,yshift=0.3cm}]{{storage/retrieval}} & & \gate[1]{\text{delay} (\tau)}
        \gategroup[1,steps=1,style={dashed,rounded corners,fill=red!20, inner xsep=2pt},background,label style={label position=above,anchor=north,yshift=0.3cm}]{{idling}} 
        & & \gate[3]{\text{retrieval}}\gategroup[3,steps=1,style={dashed,rounded corners,fill=green!20, inner xsep=2pt},background,label style={label position=above,anchor=north,yshift=0.3cm}]{}  &  \wireoverride{n} &\wireoverride{n} & \gate[3]{\text{storage}}\gategroup[3,steps=1,style={dashed,rounded corners,fill=green!20, inner xsep=2pt},background,label style={label position=above,anchor=north,yshift=0.3cm}]{} & \gate[1]{\text{delay} (\tau)}\gategroup[1,steps=1,style={dashed,rounded corners,fill=red!20, inner xsep=2pt},background,label style={label position=above,anchor=north,yshift=0.3cm}]{{}} &  \ \ldots \push{} & \gate[4]{\text{retrieval}}\gategroup[4,steps=1,style={dashed,rounded corners,fill=green!20, inner xsep=2pt},background,label style={label position=above,anchor=north,yshift=0.3cm}]{}\\
        \lstick[1,label style={blue}]{$q_s: {|0\rangle}$ }
        & \gate[2]{\text{probe-prep}}
        \gategroup[2,steps=2,style={dashed,rounded corners,fill=blue!20, inner xsep=2pt},background,label style={label position=above,anchor=north,yshift=0.3cm}]{{single sensing step}} 
        & \gate[1]{\text{sensing}}
        &  &  \wireoverride{n} &\wireoverride{n}  &  & & \gate[2]{\text{probe-prep}}\gategroup[2,steps=2,style={dashed,rounded corners,fill=blue!20, inner xsep=2pt},background,label style={label position=above,anchor=north,yshift=0.3cm}]{{}}  & \gate[1]{\text{sensing}} &  &  \wireoverride{n} & \ \ldots \\
        \lstick[1,label style={blue}]{$q_s: {|0\rangle}$ } 
        & & \gate[1]{\text{sensing}} & &   \wireoverride{n} &\wireoverride{n} & & & & \gate[1]{\text{sensing}} &  &  \wireoverride{n} & \ \ldots \\
       \lstick[1,label style={violet}]{$q_c: {|0\rangle}$ }  &&&&&&&&&&&& \ \dots&&&\gate[1]{\text{processing}}\gategroup[1,steps=1,style={dashed,rounded corners,fill=violet!20, inner xsep=2pt},background,label style={label position=below,anchor=north,yshift=-0.2cm}]{}
    \end{quantikz}}
        
\caption{Pipeline for a sensing scheme using STQS. The sensing step, storage, idling, and retrieval processes with a memory qubit are highlighted in yellow. The framework allows for an arbitrary number of steps to be implemented, showcasing its scalability and versatility. Post-processing with a computing qubit is illustrated, emphasizing the scheme's computing capabilities.} 
\label{fig:stqs-circuit}
\end{figure*}

STQS is comprised of two main components: the Quantum Sensing Processing Unit (QSPU) and the Quantum Sensing Chips (QSC). A schematic overview of the system architecture is illustrated in Figure~\ref{fig:QSS}. The QSCs consist of quantum sensors to capture physical signals, quantum memory to store quantum information and enable spatial-temporal correlated measurements, and quantum buffers to enable efficient quantum communication between QSCs, and between QSCs and the QSPU. The QSPU handles the post-processing of quantum states produced by the sensors during the sensing process and interfaces with other quantum computing components for advanced quantum processing of the acquired signals.

The STQS framework has four major stages in its pipeline, as illustrated in Figure~\ref{fig:stqs-circuit}. In the probe-preparation stage, quantum sensors within a single QSC (or across multiple QSCs) are prepared in an entangled state prior to sensing. Sensors interact with a physical field during the sensing stage to acquire signals. Results from a time slice or spatial location can be stored and later retrieved using quantum memory in the storage/retrieval stage to enable spatiotemporal correlated sensing. The quantum sensing process can be repeated multiple times, allowing for the computation of proper time correlations by adjusting the waiting time, $\tau$, between consecutive sensing processes. Finally, post-processing occurs after a communication buffer transfers data from the QSC to a computing qubit. This abstraction is central to our simulations, as detailed in Section~\ref{sec:applications}.

\begin{figure}[h]
    \centering
    \begin{tikzpicture}[node distance=1.1cm]

\node (prep) [prep_sense] {Probe Preparation};
    \node (sense) [prep_sense, below of=prep] {Sensing};
    \node (storage) [storage_retrieval, below of=sense] {Storage};
    \node (memory) [memory_style, below of=storage] {Memory};
    \node (retrieval) [storage_retrieval, below of=memory] {Retrieval};
    \node (processing) [processing, below of=retrieval] {Processing};

    \node[draw=black, dashed, fit=(storage) (memory) (retrieval), inner sep=5pt] {};

    \draw [arrow] (prep) -- (sense);
    \draw [arrow] (sense) -- (storage);
    \draw [arrow] (storage) -- (memory);
    \draw [arrow] (memory) -- (retrieval);
    \draw [arrow] (retrieval) -- (processing);

    \end{tikzpicture}
    \caption{Flow of quantum information through the components of a single iteration of an STQS sensing scheme. The sensing process progresses through probe preparation (blue), sensing (blue), storage (green), memory (red), retrieval (green), and processing (violet). The steps enclosed in the dashed box may be skipped depending on the application scenarios (for immediate quantum information processing). However, when used, they offer a decisive advantage by enabling temporal correlations across multiple iterations, a feature crucial for advanced distributed sensing protocols.}

    \label{fig:stqs-flow}
\end{figure}
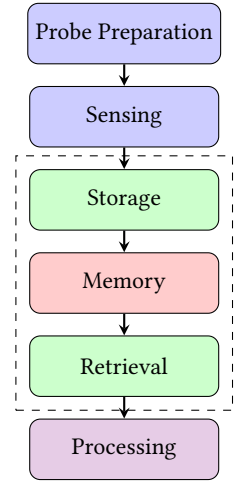

The flowchart in Figure \ref{fig:stqs-flow} provides a high-level abstraction of how quantum information moves through the key components of a single cycle of an STQS sensing process. Unlike the detailed circuit diagram in Figure \ref{fig:stqs-circuit}, the flowchart emphasizes the broader sequence of functional stages (probe preparation, sensing, storage, memory, retrieval, and processing). While sensing is essential for every iteration, the steps enclosed in the dashed box (storage, memory, and retrieval) are optional. These optional stages allow STQS to preserve quantum states for later retrieval, enabling the capture of time-separated correlations across multiple sensing events. 

The quantum information flow begins with probe preparation: initializing and entangling quantum sensors. During the sensing stage, the sensing nodes interact with external parameters to acquire signals that are then passed to a memory node for secure temporary storage. Once retrieved, the quantum data is processed using quantum algorithms that extract the relevant data. By structuring the sensing process in this modular fashion, the STQS framework remains highly adaptable, able to adapt to different time intervals, spatial configurations, and application-specific requirements. This adaptability enhances its utility across a broad range of quantum sensing scenarios.

Each stage of our design works in concert with the others to fully exploit the extreme sensitivity of quantum sensors. While the minutia of each step depends on the quantum system used as a sensor and the sensing application, the architecture is designed to facilitate robust sensing and allow for detailed studies of error propagation through the circuit. By decoupling the specifics of state preparation, phase encoding, and memory usage, we stress that STQS can be adapted to various quantum sensing applications.

\subsection{Spatially Distributed Sensing}\label{sec:spatially-distributed}

Quantum communication is a necessity in distributed sensing networks. STQS establishes a coherent measurement network that overcomes the SQL by connecting individual sensing nodes. The communication buffering of STQS allows one to temporarily store the sensed quantum information at a particular node to synchronize operations and analysis to avoid destructive communication delays. This is essential to sensing protocols that rely on entanglement-assisted measurements or non-local operations \cite{Giovannetti_2006, PhysRevLett.102.100401}.
STQS's implementation of quantum communication buffering enables joint measurements with higher spatial resolution while supporting state-of-the-art error correction by correlating data from spatially unique nodes. 

Concretely, we illustrate this by discussing the evolution of an arbitrary quantum state $|\psi\rangle$ in an STQS system. First, we establish our network of sensing qubits, initialized in the ground state, $|0\rangle$. Then, by creating a superposition of the sensing network (by entangling sensing nodes), we create the basis for correlated measurements across the distributed network, and entanglement-enabled phase accumulation. 
The STQS system is now,
\begin{equation}
    |\psi\rangle \rightarrow \frac{1}{\sqrt{2}}(|000\rangle + |111\rangle).
\end{equation}

By abstracting the sensing of spatially unique parameters as phase shifts on each sensing node, we can model the processes of quantum information encoding. The communication buffer is then introduced to temporarily store the parameter information at each node before final synchronization on a separable memory qubit node. Consequently, the state is now:
\begin{equation}
    |\psi\rangle \rightarrow \frac{1}{\sqrt{2}}(|000\rangle + e^{i(\phi_1 + \phi_2 + \phi_3)}|111\rangle).
\end{equation}
Local phases, $\phi_1, \phi_2, \text{and } \phi_3$, are accumulated across the network, creating a single, \textit{global} combined phase, $\phi_{\text{global}} = \phi_1 + \phi_2 + \phi_3$. The ability to synchronize measurements and aggregate spatially distributed data into coherent results demonstrates a significant advancement in quantum sensing.

\begin{figure}
    \centering
    \includegraphics[width=1\linewidth]{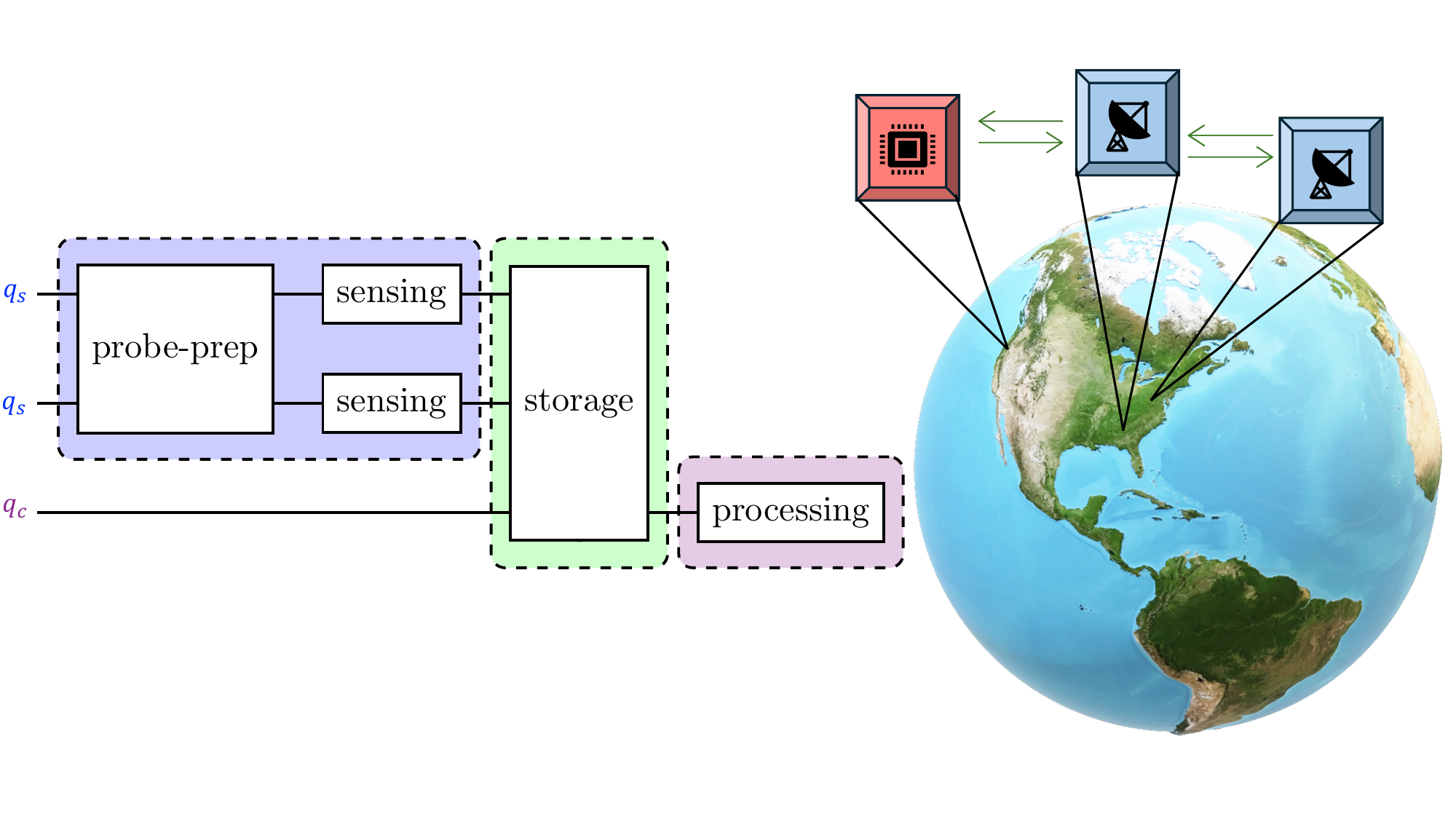}
    \caption{Schematic representation of a possible STQS spatial configuration and a circuit-level representation. The right side shows spatially distributed sensing qubits (blue) measuring local parameters, while the central processing qubit (red) buffers communication for synchronized measurements. Communication links, indicated by green arrows, connect a QSPU in Richland, WA, to distributed quantum sensors in Athens, GA, and New York City. The left side depicts the circuit-level representation, where entanglement is established among the sensing qubits before being transferred to the processing qubit for coherent aggregation.}
    \label{fig:spatial-correlations}
\end{figure}

Figure \ref{fig:spatial-correlations} illustrates the spatial separation of the sensing qubits (blue) and the central quantum processing unit (red) to highlight how each sensor can be physically located at distinct nodes. Some communication links are indicated by green arrows. By placing the sensing qubits at different positions, the network can acquire locally unique phase information while maintaining coherence through a shared quantum channel. The circuit-level diagram on the left side of Figure \ref{fig:spatial-correlations} shows how entanglement is generated among the sensing qubits and subsequently transferred to the processing qubit, which functions as a communication buffer. STQS's arrangement ensures that each sensor's measurements can be collected and synchronized, enabling global phase estimation across widely separated nodes. Crucially, the processing qubit not only aggregates the information but also serves as a memory resource that can be isolated from local noise, possibly enhancing the overall fidelity of the sensing protocol.

\subsection{Temporally Correlated Sensing}

Determining temporal correlations from quantum sensors is critically important, as real-world signals outside the laboratory are rarely composed of static, unchanging parameters. 
These time-dependent fluctuations are often present when detecting weak or fluctuating signals like those found in gravitational waves when distant astrophysically large-scale events (merging black holes or neutron stars) emit faint and time-varying gravitational fluctuations \cite{Will_2014, Abbott_2016}. Temporally correlated sensing allows us to track changes and build sensitivity over time to amplify the ability to detect weak signals even when noise is present. A more terrestrially relevant example is environmental monitoring. Temporal fluctuations in temperature, magnetic fields, and pollutant concentrations are critical for accurate assessments and timely interventions.
Climate science will benefit \cite{Nammouchi2024QuantumML} from sensing systems using STQS for time-correlated sensing, as will security and military applications where time correlations of magnetic and temperature parameters are indicators of emerging anomalies and threats \cite{10.1117/12.441262}. Moreover, these same principles are closely aligned with robust error correction in quantum systems, non-Markovian noise \cite{Breuer_2009, RevModPhys.89.015001} exhibits structured time-dependent behavior rather than purely random, memory-free fluctuations.

The memory qubit of our model is the abstraction of these described time-dependent fluctuations in sensed parameters. STQS can accumulate phase information by applying a delay gate (parameterized by time duration $\tau$), integrating contributions from temporal variations in the sensed parameters. By considering the most general and arbitrary initial state of the memory qubit as
\begin{equation}
    |\psi\rangle = \alpha|0\rangle + \beta|1\rangle,
\end{equation}
where $\beta = e^{i \sum^{N}_{i=1}\phi_i}$ represents the unknown parameters accumulated at the initial time across the sensing qubits, we see that if the qubit undergoes the action of the delay gate for time $\tau$:
\begin{equation}
    |\psi\rangle \rightarrow  \alpha|0\rangle + \beta e^{i\tau}|1\rangle =  \alpha|0\rangle + e^{i \sum^{N}_{i=1}\phi_i + \tau}|1\rangle.
\end{equation}

The delay gate updates the phase term to include the delay-induced shift $\tau$. The memory qubit functions as a robust recorder of temporal dynamics, capable of capturing both initial fluctuations of the sensed parameter and subsequent evolution over time. 

\begin{figure}
    \centering
    \includegraphics[width=1\linewidth]{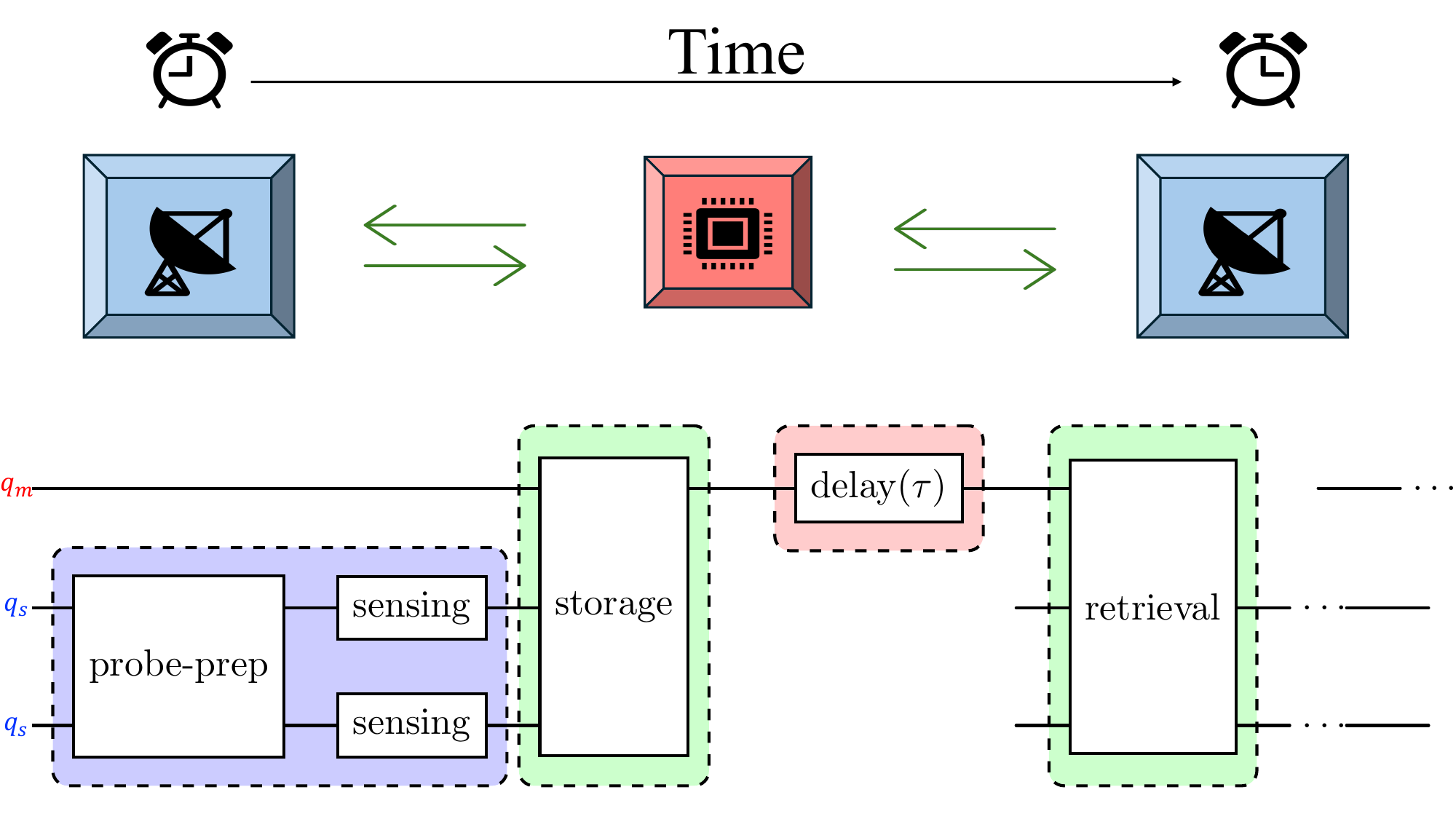}
    \caption{Schematic (top) and circuit-level (bottom) representations of the STQS architecture for temporally correlated sensing. The memory qubit (red) integrates unknown parameters over a time-delay interval $\tau$, and distributed sensing qubits (blue) sample the local unknown parameters. Green arrows denote communication links for data exchange. In the circuit diagram, a delay gate of duration $\tau$ captures the temporal evolution of the sensed parameters, enabling synchronized measurements over extended timescales.}
    \label{fig:time-correlations}
\end{figure}

Figure \ref{fig:time-correlations} illustrates the core elements of the STQS architecture for time-correlated quantum sensing. In the top schematic, the memory qubit (red) is responsible for housing the aggregated sensed parameters over an interval $\tau$, while sensor nodes (blue) are linked via communication channels (shown with green arrows), ensuring coherent data transfer across the network. The bottom circuit-level diagram highlights how a delay operation of duration $\tau$ is applied to the memory qubit; although this circuit diagram doesn't specify exact operations, it clarifies the circuit-level building blocks involved in capturing and aggregating time-varying signals.

\subsection{State Preparation}\label{sec:state-prep}

Within the current landscape of quantum sensing, a crucial gap persists between the need and ability to develop efficient microarchitectures for quantum state preparation. We have designed STQS to help bridge that gap. A key component of STQS's design is quantum state preparation. The initial quantum state of a sensing system determines the sensitivity, accuracy, and robustness of the measurements. The quantifier for how much information a quantum state can carry about an unknown parameter is the quantum Fisher information (QFI), which directly influences the sensitivity of measurements. This is because the QFI, $F(\Theta)$ of an unknown parameter $\Theta$ (with unbiased estimator $\hat{\Theta}$), is related to the precision of parameter estimation over $N$ measurements through the Quantum Cramér-Rao bound given by, 
\begin{equation}
    \text{Var}(\hat{\Theta}) \geq \frac{1}{N*F(\Theta)}.
\end{equation}
Optimizing the probe state of a quantum sensing system to have the greatest possible QFI also optimizes the variance in parameter estimation and sensitivity. By optimizing the probe states' QFI, a sensitivity that overcomes the SQL is achieved to approach or reach the HL under noiseless conditions. It has been previously shown \cite{vasilyev2024optimalmultiparametermetrologyquantum} that in any noiseless discrete variable system, the QFI will always be maximized by preparing the probe states of a quantum sensor in the GHZ state. Similar results have been shown in continuous variable setups, two-mode squeezed vacuum states (TMSV) \cite{vasilyev2024optimalmultiparametermetrologyquantum, PhysRevA.72.032334} and more recently non-Gaussian GKP states have been shown to be similarly promising \cite{Hanamura_2021, PhysRevA.95.012305}. In the following section, \ref{sec:QML}, we will explore how quantum machine learning techniques can be leveraged to enhance state preparation for quantum sensing applications where the effects of noise are relevant.

\subsection{Quantum Processing and Machine Learning}\label{sec:QML}

Processing via QSPUs is a critical resource within the STQS framework. The QSPU computing nodes facilitate the implementation of advanced quantum machine learning algorithms and serve as versatile processors for real-time quantum data analysis and feedback control. While our focus in the following discussion is on quantum machine learning (QML) due to its potential for adaptive optimization of sensing processes, it is essential to note that the STQS architecture is not limited to QML techniques alone. Other quantum algorithms can also be deployed on these computing nodes, such as quantum error correction, variational quantum eigensolvers, and specialized signal processing routines.

Quantum machine learning offers an effective approach to enhancing quantum sensing by enabling adaptive and intelligent control over state preparation and subsequent evolution post-sensing \cite{maclellan2024endtoendvariationalquantumsensing}. In the STQS framework, QML methods are mapped onto key functional modules to refine sensing performance and mitigate the effects of noise through real-time feedback and iterative optimization. Furthermore, by efficiently encoding information, QML-enabled quantum sensing has potential applications for system-resource overhead reduction. In the following subsections, we outline using the swap test as a tool for quantifying the fidelity between a sensed quantum state and a user-provided reference state. This example showcases the streamlined integration of QML techniques into a sensing protocol within the STQS architecture.

\subsubsection{QML-Enhanced Adaptive Quantum Sensing}
\begin{figure}
    \centering
    \includegraphics[width=\linewidth]{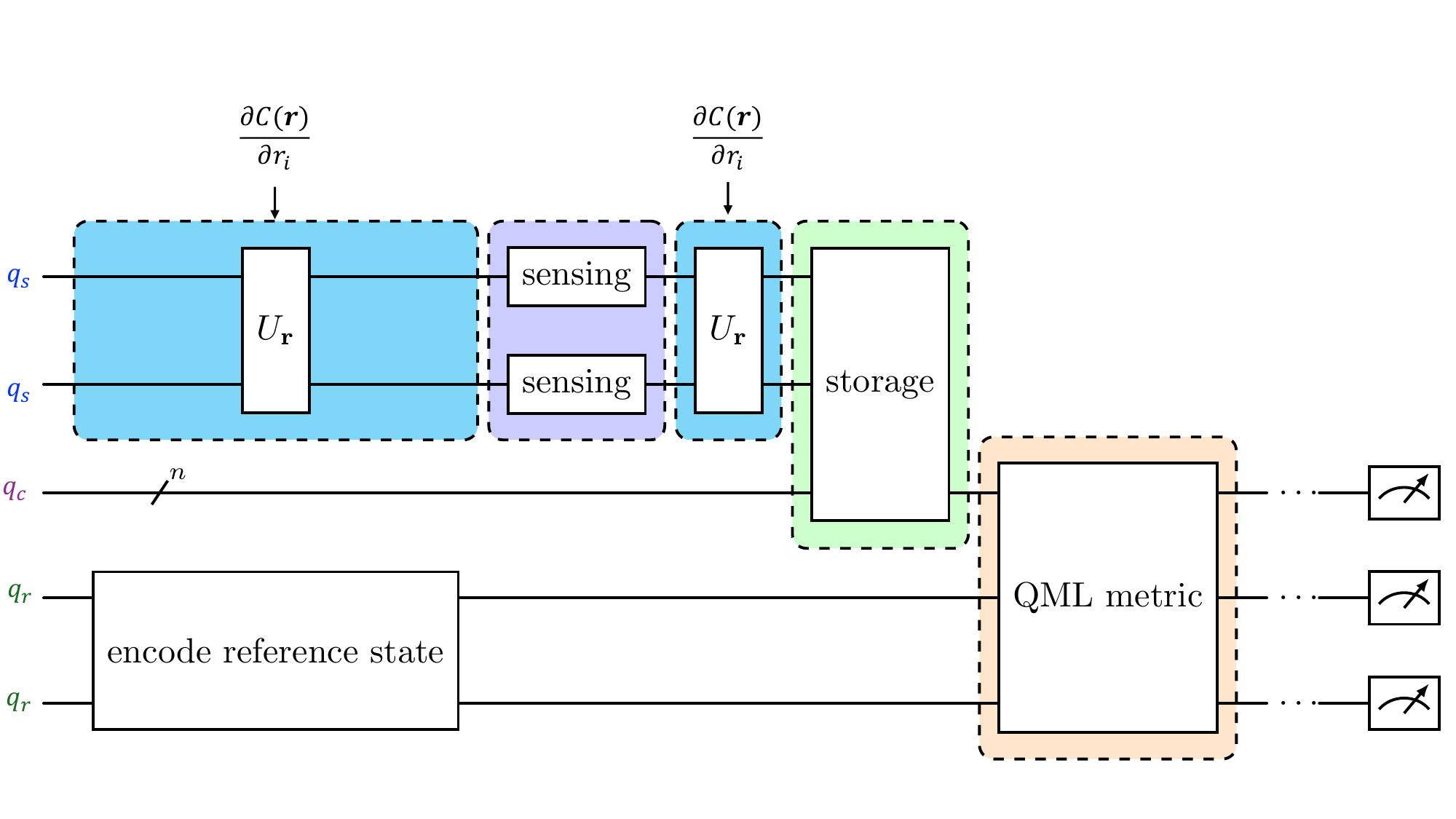}
    \caption{QML-enhanced adaptive quantum sensing within the STQS architecture. The figure features four color-coded components: parameterized unitaries (cyan,) the sensing process (purple), a storage/communication interface (green) for state transfer to a computing node, and a user-defined QML metric (orange) that informs an iterative gradient-descent optimization of the variational parameters.}
    \label{fig:qml-diagram}
\end{figure}

The general\\ scheme for quantum sensing enhanced by QML is shown in Figure \ref{fig:qml-diagram} and illustrates how key components within the STQS architecture work together to optimize a distributed sensing process. The diagram is partitioned into four key sections, each identified by a distinct color: the parameterized unitaries in cyan, the sensing process in purple, the storage/communication interface in green, and a user-provided QML metric in orange. The cyan boxes represent the parameterized unitaries $U(\mathbf{r})$, where $\mathbf{r}$ is a vector of variational parameters. The purple box illustrates the sensing process during which the quantum system interacts with its environment to encode information into the quantum state. The green box indicates the communication stage, where the sensed state is transferred to the computing nodes, enabling optimized processing. The orange box shows the user-provided QML metric, which is essential for evaluating the performance of the quantum sensing process.

Mathematically, QML utilizes the output of a quantum algorithm to evaluate a desired metric. Suppose the output state from the sensing process is $\rho_{\text{sens}}$, and the reference state is $\rho_{\text{rho}}$. A fidelity metric $M$ or user-specified function $f(\rho_{\text{sens}}, \rho_{\text{rho}})$ is then computed to quantify the similarity between the states. This metric is incorporated into a cost function $C(\mathbf{r})$ to measure the deviation of performance from an ideal outcome. The optimization process is iterative and typically uses gradient descent \cite{maclellan2024endtoendvariationalquantumsensing}---a method where the gradient of the cost function with respect to the variational parameters \(\mathbf{r}\) is computed, and the parameters are updated accordingly. Specifically, for the cost function \( C(\mathbf{r}) \), the gradient with respect to each parameter \(r_i\) is given by:
\begin{equation}
    \frac{\partial C(\mathbf{r})}{\partial r_i}.
\end{equation}
In practice, each parameter is updated iteratively as:
\begin{equation}
    r_i^{(t+1)} = r_i^{(t)} - \eta \frac{\partial C(\mathbf{r}^{(t)})}{\partial r_i},
\end{equation}
where \(\eta\) is the learning rate and \(t\) is the iteration step. The results of the update rule are fed back into $U(\mathbf{r})$ accordingly. Integrating measured outcomes from the sensing process with a user-provided metric within the STQS framework, QML-enhanced adaptive quantum sensing will likely enhance sensitivity and reduce overhead.

\subsubsection{Prediction of Quantum Sensing}\label{sec:SWAP}

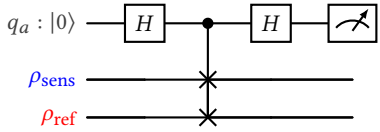
\begin{figure}
    \centering
    \begin{quantikz}
        \lstick[label style={darkgray}]{$q_a: |0\rangle$} & \gate{H} & \ctrl{2} & \gate{H} & \meter{} \\
        \lstick[label style={blue}]{$\rho_{\text{sens}}$} & \qw & \swap{1} & \qw & \qw \\
        \lstick[label style={red}]{$\rho_{\text{ref}}$} & \qw & \targX{} & \qw & \qw
    \end{quantikz}
    \caption{Circuit diagram of the swap test. The uppermost qubit is an ancilla computing qubit.}
    \label{fig:SWAP-circuit}
\end{figure}

In the future, we anticipate scenarios where multiple users can access a distributed quantum sensing system simultaneously. During such interactions, users can provide a reference quantum state for comparison and verification in particular sensing applications. This reference state can either be directly input through quantum teleportation via a quantum network, which can theoretically ensure privacy and security since the reference quantum state is unknown and measurement can destroy the state, or online generated based on a classical description provided by the users, through quantum encoding. In these cases, it becomes essential to establish a robust method to measure the ``distance'' between the sensed quantum state and the user-provided reference state. This enables evaluation of how closely the sensed state aligns with the user’s desired state to predict whether an event has occurred or a metric of interest has reached a certain degree.

There are various methods to characterize and measure quantum states. We propose using the swap test algorithm as a practical approach for such a distance measurement task. The swap test compares the reference and sensed states by determining the overlap or fidelity between them. Enabling quantification of the sensed state's confidence level in terms of how it matches the reference state specified by the user. Moreover, the swap test can be integrated with adaptive feedback mechanisms to refine the sensing process further. For instance, by continuously monitoring the fidelity and applying real-time adjustments, the system can compensate for environmental noise or dynamic changes in the signal.

While the swap test mentioned in this section is our canonical example of a possible choice of how to measure state overlap or distance, we wish to stress that other approaches (like state tomography and interference-based methods) may also be leveraged, depending on the specific requirements and constraints of the sensing application. Beyond the fundamental utility in quantum state comparison, the swap test has also previously been used for QML ~\cite{Stein_2021}. In fact, by leveraging the existing body of QML research that integrates swap tests, we can further extend the capability of STQS for more advanced sensing data processing and learning tasks. 

This dual advantage makes the swap test an appealing choice: it provides a precise metric for state similarity and enables advanced sensing state analysis through QML techniques. By embedding this capability, the system becomes a tool for accurate state sensing and a platform for adaptive and intelligent quantum applications. This integration could significantly enhance the flexibility and efficiency of quantum sensing, allowing it to serve a broader range of use cases where state-specific precision and learning-based adaptability are critical.

\section{Applications and Evaluations} \label{sec:applications}
Here, we validate the usefulness of STQS in current sensing endeavors, specifically quantum radar and dark matter detection. We present a simulation of quantum radar-based remote sensing of soil saturation using the characteristics of Rydberg atom qubits from Table \ref{tab:qubit_errors2} inspired by the ground-based proof of concept work in \cite{arumugam2024remotesensingsoilmoisture}. Additionally, we present a simulation of wavelike dark matter detection using the characteristics of transmon-based superconducting qubits from Table \ref{tab:qubit_errors2}. In both simulations, we explore the key parameter space domains that may prove useful to future experimental designs.

We engineered a custom noisy quantum simulator specifically designed to meet the requirements of quantum sensing applications, implementing highly precise and streamlined customization of noise profiles. By tailoring the simulator to accurately model error profiles across diverse quantum systems and fine-tuning specific noise types on modular qubits, we ensured flexibility and control unmatched by existing platforms. This custom approach allowed us to explore the nuanced effects of various noise parameters, providing critical insights into their impact on the performance of quantum sensors.

\subsection{Quantum Radar}\label{sec:quantum radar}

\begin{figure}
    \centering
    \includegraphics[width=\linewidth]{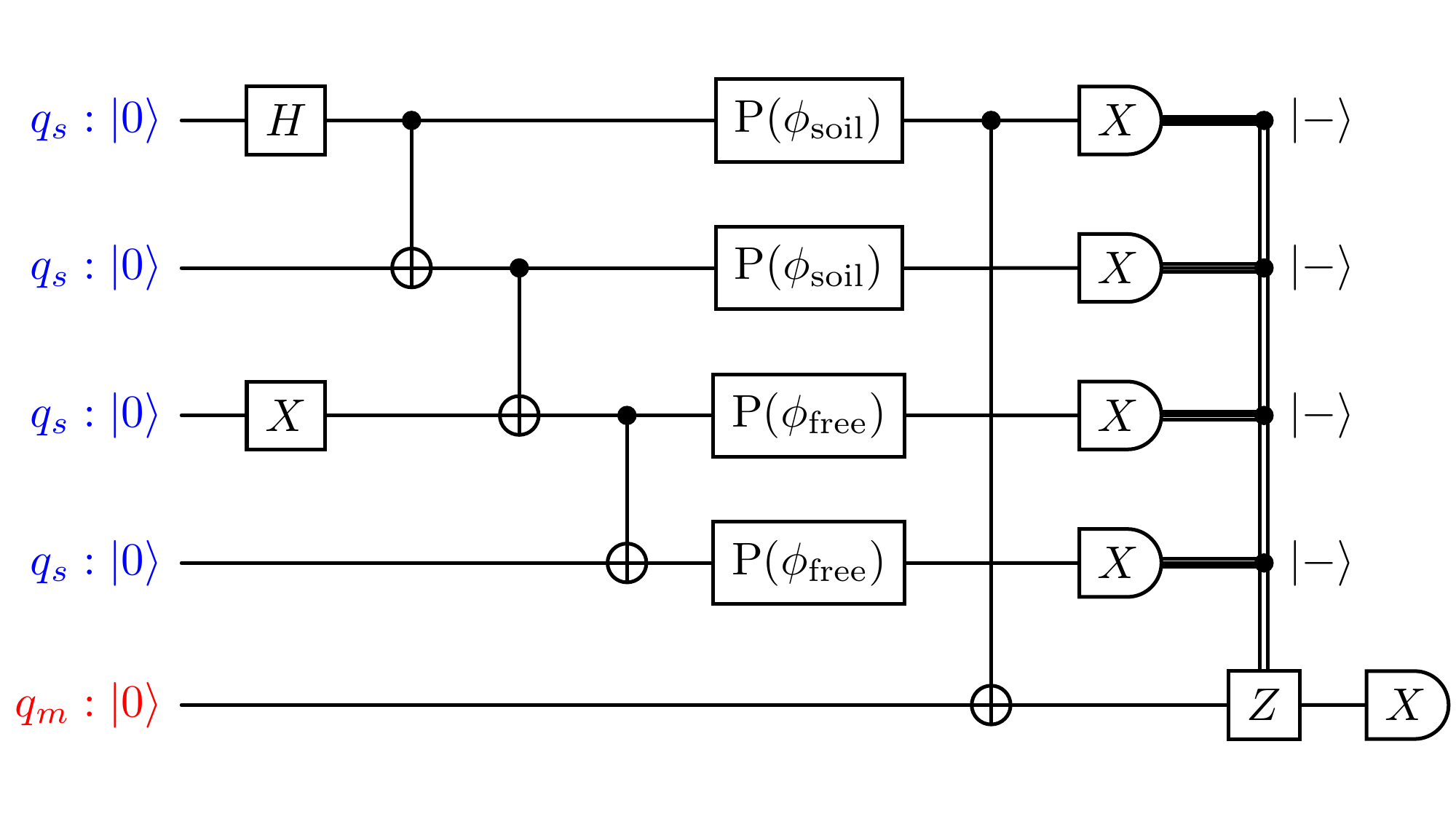}
    \caption{Diagram of the quantum circuit designed to compute the phase difference between the free space signal (\(\phi_\text{free}\)) and the soil signal (\(\phi_\text{soil}\)) after one sensing step. After measurement in the $x$-basis, a $Z$-gate is applied to the memory qubit for each sensing qubit measurement outcome of $|-\rangle$, while no $Z$-gate is applied for a $|+\rangle$ measurement. See Section \ref{sec:quantum radar} for a discussion of an alternative method to implement this correction in post-processing without applying physical $Z$-gates.}
    \label{fig:radar-circuit}
\end{figure}

Quantum radar is an application of quantum sensing that relies on the entanglement of quantum systems to surpass the limits of classical radar systems and is particularly useful for targeting low-reflectivity and distant targets. By entangling pairs of quantum systems at a source point and sending only one of the systems that make up a pair to the target while retaining the other, quantum radar protocols can detect changes in the returned signal with a precision exceeding what is classically possible. To date, most experimental demonstrations of quantum radar and theoretical literature surrounding the protocol have focused on CV quantum systems \cite{doi:10.1126/science.1160627, Shapiro_2009, Zhang_2013} where entanglement is used to generate correlation in squeezed quadratures of light. In this work, we present a mapping of CV-based quantum radar experiments to DV-based simulations in the circuit diagram in Figure \ref{fig:radar-circuit} using STQS. 

The recent publication \cite{arumugam2024remotesensingsoilmoisture} presents the innovative technique of using classical signals of opportunity (SoOp) of XM-satellites to remotely measure soil moisture. Extremely compact, sensitive, and broad spectrum (micro to millimeter frequencies) Rydberg atoms act as quantum probes capable of detecting weak electromagnetic fields without bulky and impractical RF band-specific electronic receivers. 
The experimental work of \cite{arumugam2024remotesensingsoilmoisture} has demonstrated that Rydberg atoms can correlate reflected signals to measure soil moisture with high precision, with the sensing bandwidth corresponding to target soil depth. However, these recent experiments have been ground-based and limited to unentangled Rydberg atom probes. By not employing entanglement, these experiments have not yet taken full advantage of the unique properties of quantum systems, particularly the enhanced sensitivity and noise resilience that entangled states provide.

Our STQS based simulations incorporate entangled probes and represent a novel study in this domain. Our work shows that substantial quantum enhancements will further improve the precision and scope of these experiments, potentially opening new avenues for applications in remote quantum radar. Following the results of \cite{arumugam2024remotesensingsoilmoisture} and with a desire to better understand the impact of noise on this application, as well as to quantify the potential benefits of noise reduction and the scalability advantages of using squeezed or entangled quantum resources rather than purely classically correlated Rydberg atoms we begin our investigation of quantum radar as a means of exploring the potential enhancements to remote soil moisture sensing. Specifically, we examine the scenario where probes sense signals corresponding to free space $\phi_\text{free}$ and compare it to the signal coming from some target soil sample, $\phi_\text{soil}$. The relationship between $\phi_\text{free}$ and $\phi_\text{soil}$ is key to determining the target soil's saturation. As is using the relationship, $\phi_\text{soil} = \frac{\phi_\text{free}}{\sqrt{\varepsilon_\text{soil}}}$, where \( \varepsilon_\text{soil} \) is the dielectric permittivity of the soil, a known parameter relating to the soil's volumetric water content via the Topp equation \cite{1980WRR....16..574T}. The empirical Topp equation links the dielectric permittivity and water content in mineral soils, making it an essential tool for interpreting the data.

Determining the relationship between $\phi_\text{free}$ and $\phi_\text{soil}$ is crucial to determining the soil's saturation, and thus, directly motivated the design of our circuit in Figure \ref{fig:radar-circuit}. By encoding $\phi_\text{free}$ and $\phi_\text{soil}$  into the quantum states and exploiting their interference, the circuit effectively extracts the phase difference, providing the key to assessing the soil's dielectric properties. We prepare the probe states as GHZ-type states, as prescribed following the discussion in Section \ref{sec:state-prep}, with a bit flip applied to the qubits that accrue the free space signal $\phi_{\text{free}}$ to ensure that the different signals will interfere. 

To fully understand how Figure \ref{fig:radar-circuit} incorporates STQS for quantum radar, let's discuss a detailed circuit-level breakdown and highlight how Figure \ref{fig:radar-circuit} implements signal buffering and phase encoding to enhance sensing. The qubits in Figure \ref{fig:radar-circuit} labeled $q_s$ act as sensing nodes, with separate ensembles dedicated to free space or soil signal acquisition. The number of sensors dedicated to each medium is denoted as $n_s$ for soil and $n_f$ for free space. The probe state preparation process commences by entangling all sensing qubits into a GHZ state. However, a deliberate $X$ gate is applied to the first qubit in the free space sensing ensemble to introduce a relative phase shift. This ensures that the accumulated phase in the sensing ensemble explicitly encodes the difference between the free space and soil signals rather than their cumulative absolute value. Signal acquisition is subsequently modeled through phase operators $\text{P}(\phi_{\text{soil}})$ or $\text{P}(\phi_{\text{free}})$, applied to the respective sensing modality qubits.  A CNOT operation is employed to facilitate information transfer, mapping the acquired phase information from sensing qubits $q_s$ to the memory qubit $q_m$. Due to the nature of the GHZ state, information about the acquired phases is inherently shared across the entire ensemble, meaning that the CNOT operation could, in principle, be applied to any of the sensing qubits. In our circuit representation, we arbitrarily depict the CNOT acting on the first qubit of the sensing ensemble, but the operation would have the same effect regardless of which qubit is chosen. 

Classically controlled operations are implemented based on the measurement outcomes after the sensing qubits are measured in the $x$-basis. If any sensing qubit measurement yields a $|-\rangle$, a $Z$ gate is applied to the memory qubit to correct for the introduced phase flip. However, the $Z$ gates do not need to be physically applied in a practical implementation. Instead, the correction can be handled with post-processing by keeping track of all measurement outcomes and applying a phase flip to the final outcome only if an odd number of $|-\rangle$ measurements are found. Note that the post-processing approach reduces circuit depth and mitigates the need for additional quantum gates, making the implementation more resource-efficient.

The communication buffering in our mapping to the STQS architecture provides a substantial practical advantage. Specifically, the time required to acquire signals varies significantly depending on whether the source comes from free space, sensed soil, or distinct satellite SoOps. By incorporating this buffering mechanism, all incoming data streams can be synchronized, ensuring that measurements from disparate sources are aligned prior to post-processing and measurement. This temporal correlation aids in maintaining the fidelity of the sensing protocol, especially for remote sensing schemes where real-time processing is infeasible and discrepancies in acquisition timing could introduce systematic errors in data interpretation.

Our first simulation of quantum radar analyzes the accuracy in sensing for the noise profiles of the quantum systems in Table \ref{tab:qubit_errors2}. While all these qubit types (aside from Rydberg atoms) have not yet been studied in a soil sensing experimental design, studying their noise profiles in this scheme provides an essential perspective on the robustness and the practicality of using different quantum systems. Additionally, we stress that these noise profiles correspond to already physically achievable levels in current experimental setups, making them relevant to understanding how the circuit will perform in a broader range of applications. 

\begin{figure}
    \centering
    \begin{tikzpicture}
    \begin{axis}[
        width=\columnwidth, 
        height=0.7\columnwidth, 
        xlabel={Number of Shots},
        ylabel={Accuracy (\%)},
        xmode=log,
        grid=both,
        legend style={
            at={(0.5,-0.25)}, 
            anchor=north,
            legend columns=3, 
            font=\small
        },
        legend cell align={left},
        tick label style={font=\footnotesize},
        label style={font=\small},
        title style={font=\small}
    ]

    \addplot[
        color=red,
        mark=*,
        mark options={solid, fill=red},
        style=solid
    ] coordinates {
        (100, 98.9416)
        (1000, 99.4654)
        (10000, 99.8389)
        (100000, 99.9959)
    };
    \addlegendentry{Noiseless}

    \addplot[
        color=BurntOrange,
        mark=square*,
        mark options={solid, fill=BurntOrange},
        style=solid
    ] coordinates {
        (100, 80.7943)
        (1000, 82.8695)
        (10000, 81.2240)
        (100000, 80.4546)
    };
    \addlegendentry{Superconducting}

    \addplot[
        color=ForestGreen,
        mark=triangle*,
        mark options={solid, fill=ForestGreen},
        style=solid
    ] coordinates {
        (100, 74.6921)
        (1000, 77.3614)
        (10000, 75.6858)
        (100000, 75.7417)
    };
    \addlegendentry{NV center}

    \addplot[
        color=violet,
        mark=diamond*,
        mark options={solid, fill=violet},
        style=solid
    ] coordinates {
        (100, 73.8398)
        (1000, 70.6298)
        (10000, 72.6946)
        (100000, 72.9715)
    };
    \addlegendentry{Rydberg atom}

    \addplot[
        color=Periwinkle,
        mark=pentagon*,
        mark options={solid, fill=Periwinkle},
        style=solid
    ] coordinates {
        (100, 91.2302)
        (1000, 92.4711)
        (10000, 92.9226)
        (100000, 92.4283)
    };
    \addlegendentry{Trapped ion}

    \end{axis}
    \end{tikzpicture}
    \caption{Measurement accuracy of the simulated quantum radar application for different qubit noise profiles as a function of the number of shots. $\phi_{\text{soil}} = 0.1$ and $\phi_{\text{free}} = 0.9$.} 
    \label{fig:radar-results}
\end{figure}
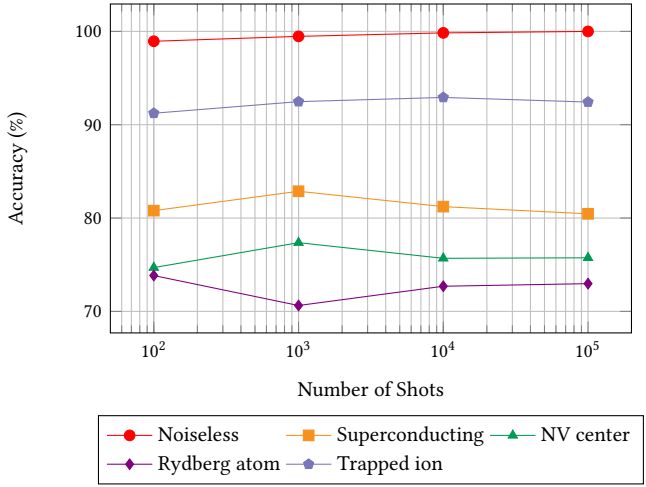

Our results for this demonstration are illustrated in Figure \ref{fig:radar-results}; for all additional demonstrations (aside from the one shown in Figure \ref{fig:radar-results} of the quantum radar application), we choose to focus on refining the noise levels specifically for the Rydberg system, as it is the platform with the most significant experimental progress and is the most promising candidate for near-term implementation. We vary the number of shots for this simulation and note that increasing the number of shots did not substantially affect the accuracy of the measurements. Based on these results, we fixed the number of shots to $10^6$ for the remainder of the experiments to streamline the analysis. Across all tested noise profiles, the noiseless results unsurprisingly had the greatest accuracy, followed by trapped ion systems, superconducting qubits, bulk NV centers, and Rydberg atom systems. For these simulations, the number of sensor probes used was three qubits for the soil signal ($\phi_{\text{soil}} = 0.9$) and three qubits for the free space signal ($\phi_{\text{free}} = 0.1$), for a phase difference in the signals of $0.8$. Our results highlight just how much the noise profile of a quantum system used for sensing impacts the accuracy of the measurement, underscoring the possible sustainability of different qubit types for quantum sensing tasks.

\begin{figure}
    \centering
    \begin{tikzpicture}
    \begin{axis}[
        width=\columnwidth,
        height=0.7\columnwidth, 
        xlabel={Number of Sensing Qubits ($n_{\text{soil}} = n_{\text{free}}$)},
        ylabel={Accuracy (\%)},
        symbolic x coords={4, 6, 8, 10, 12},
        xtick=data,
        ymin=-270, ymax=110,
        ymajorgrids=true,
        grid=both,
        bar width=8pt, 
        enlarge x limits=0.15,
        legend style={
            at={(0.7,.5)}, 
            anchor=north,
            font=\small
        },
        legend cell align={left},
        tick label style={font=\small},
        label style={font=\small},
        title style={font=\small},
        legend image code/.code={
            \draw[#1,fill=#1] (0cm,-0.1cm) rectangle (0.3cm,0.1cm);
        }
    ]

    \addplot[
        ybar,
        bar shift=-20pt,
        fill=red,
        draw=black
    ] coordinates {
        (4, 99.0775)
        (6, 99.5637)
        (8, 98.9093)
        (10, 99.94)
        (12, 99.6068)
    };
    \addlegendentry{Noiseless}

    \addplot[
        ybar,
        bar shift=-10pt,
        fill=BurntOrange,
        draw=black
    ] coordinates {
        (4, -255.6238)
        (6, -128.5835)
        (8, -56.9266)
        (10, -12.7884)
        (12, 13.6511)
    };
    \addlegendentry{Default}

    \addplot[
        ybar,
        bar shift=0pt,
        fill=ForestGreen,
        draw=black
    ] coordinates {
        (4, -197.1870)
        (6, -121.1462)
        (8, -66.9829)
        (10, -8.4074)
        (12, 15.3902)
    };
    \addlegendentry{$T_1 = \infty$}

    \addplot[
        ybar,
        bar shift=10pt,
        fill=blue,
        draw=black
    ] coordinates {
        (4, -208.6846)
        (6, -122.7916)
        (8, -46.7361)
        (10, -11.9017)
        (12, 12.1320)
    };
    \addlegendentry{$T_2 = \infty$}

    \addplot[
        ybar,
        bar shift=20pt,
        fill=violet,
        draw=black
    ] coordinates {
        (4, -192.3006)
        (6, -86.4109)
        (8, -46.6470)
        (10, -8.2824)
        (12, 15.8599)
    };
    \addlegendentry{$T_1 = T_2 = \infty$}

    \addplot[
        thick, gray, dashed,
        forget plot 
    ] coordinates {
        (4, 0) (12, 0)
    };

    \end{axis}
    \end{tikzpicture}
    \caption{Measurement accuracy of the simulated quantum radar application with a varying number of probes ($n_s = n_f$) and coherence ($T_1$ and $T_2$) times.}
    \label{fig:radar-bar-graph-t1-t2}
\end{figure}
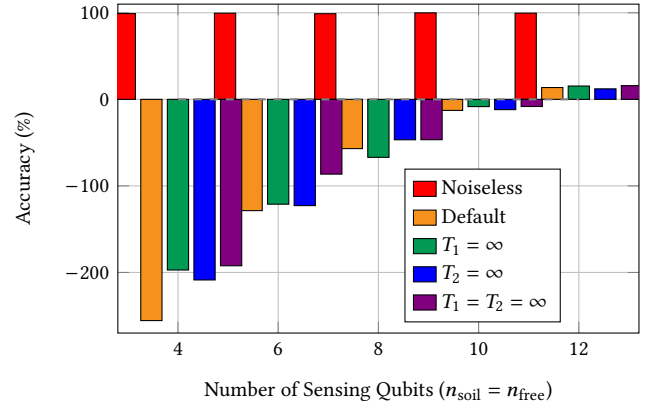

We now turn to Figure \ref{fig:radar-bar-graph-t1-t2}, which presents the results of simulations designed to study how measurement accuracy changes as a function of both coherence times ($T_1$ and $T_2$) of the Rydberg atoms used for the quantum radar application and the number of sensing probes used (where $n_f = n_s$ for all simulations). The results demonstrate that varying $T_1$ and $T_2$ had a noticeable impact on accuracy when the number of sensing qubits is small (e.g., four qubits). However, as the number of sensing qubits is increased to six, the effect of an infinite $T_1$ and $T_2$ diminishes, with accuracy showing minimal variance compared to the default noise profile. These findings suggest that for quantum radar applications, it is more beneficial to focus on generating a larger ensemble of probes rather than improving the coherence times of individual qubits. It's clear that the number of probes outweighs the benefits of increased coherence times in this regime.

\begin{figure}
    \centering
    \begin{tikzpicture}
    \begin{axis}[
        width=\linewidth,
        height=0.6\linewidth,
        xlabel={Number of Sensors},
        ylabel={Accuracy (\%)},
        symbolic x coords={4, 6, 8, 10},
        xtick=data,
        ymin=-320, ymax=5,
        ytick={-350,-300,-250,-200,-150,-100, -50, 0},
        bar width=10pt,
        enlarge x limits=0.2,
        legend style={at={(0.5,-0.3)}, anchor=north, legend columns=2},
        legend cell align={left},
        grid=both,
        legend image code/.code={
            \draw[#1, fill=#1] (0cm,-0.1cm) rectangle (0.3cm,0.1cm);
        }
    ]

    \addplot[
        ybar,
        bar shift=-15pt,
        fill=red
    ] coordinates {
        (4,-315.5125)
        (6,-188.3939)
        (8,-94.1936)
        (10,-33.7507)
    };
    \addlegendentry{Default}

    \addplot[
        ybar,
        bar shift=-5pt,
        fill=red,
        opacity=0.7
    ] coordinates {
        (4,-225.9048)
        (6,-109.6828)
        (8,-55.6430)
        (10,-35.4820)
    };
    \addlegendentry{Readout}

    \addplot[
        ybar,
        bar shift=5pt,
        fill=red,
        opacity=0.5
    ] coordinates {
        (4,-194.6076)
        (6,-134.4310)
        (8,-66.3443)
        (10,-18.1234)
    };
    \addlegendentry{Single-qubit gate}

    \addplot[
        ybar,
        bar shift=15pt,
        fill=red,
        opacity=0.3
    ] coordinates {
        (4,-196.0395)
        (6,-156.1460)
        (8,-70.5664)
        (10,0.8442)
    };
    \addlegendentry{Multi-qubit gate}

    \end{axis}
    \end{tikzpicture}
    \caption{Accuracy of the soil sensing circuit as a function of the number of sensor qubits (\(n_f = n_s\)) under different noise profile conditions. The phase difference between \(\phi_\text{soil}\) and \(\phi_\text{free}\) = 0.13 for these simulations.}

    \label{fig:grouped-bar-phi-0.1}
\end{figure}
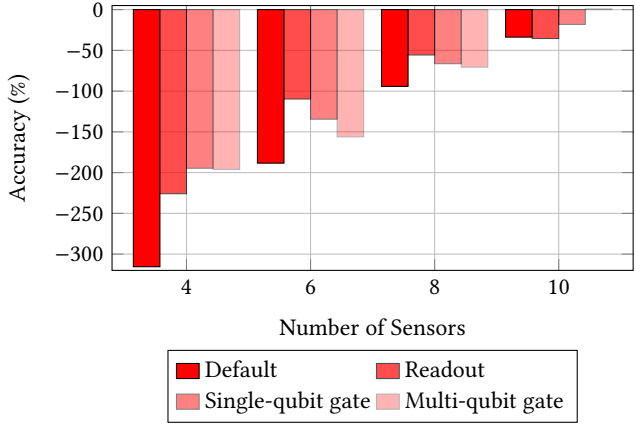

Figure \ref{fig:grouped-bar-phi-0.1} corresponds to the simulations that examine the effect of selectively removing specific noise types--multi-qubit gate errors, single-qubit gate errors, or readout errors--while keeping the remaining noise sources at their default levels. This analysis is done repeatedly with varying numbers of sensing probes ($n_f = n_s$). The results reveal a trend similar to what is observed when varying the $T_1$ and $T_2$ times: scaling the number of probe sensor qubits yields a more significant accuracy improvement than completely eliminating a single type of noise. Once again, this highlights that for distributed quantum radar, increasing the number of probes seems to be a more effective strategy for improving the performance of accuracy in measurement than only focusing on reducing specific noise contributions.

\begin{figure}
\centering
\begin{tikzpicture}
    \begin{axis}[
        width=\columnwidth,  
    height=0.7\columnwidth,
        xlabel={$\epsilon$},
        ylabel={Accuracy (\%)},
        xmin=0, xmax=1,
        ymin=-90, ymax=100,
        xtick={0,0.1,0.2,0.3,0.4,0.5,0.6,0.7,0.8,0.9,1},
        ytick={-80,-60,-40,-20,0,20,40,60,80,100},
        grid=both,
        minor grid style={gray!25},
        major grid style={gray!50},
        legend pos=south east
    ]
        \addplot[smooth,blue,thick] coordinates {
            (0, -68.4142)
            (0.1, -65.7277)
            (0.2, -58.4775)
            (0.3, -57.1679)
            (0.4, -42.3653)
            (0.5, -30.2246)
            (0.6, -13.3040)
            (0.7,2.2853)
            (0.8,29.0099)
            (0.9, 59.3554)
            (1, 99.6819)
        };
        \addlegendentry{Gate \& readout}

        \addplot[smooth,black,thick] coordinates {
            (0, 2.6768)
            (0.1, 17.6315)
            (0.2, 28.7477)
            (0.3, 32.8237)
            (0.4, 42.1574)
            (0.5, 42.8574)
            (0.6, 57.9154)
            (0.7, 61.8342)
            (0.8, 67.2060)
            (0.9, 87.9612)
            (1,  99.6819)
        };
        \addlegendentry{Gate}

        \addplot[smooth,red,thick] coordinates {
            (0,-53.6449)
            (0.1, -45.6055)
            (0.2,  -45.0918)
            (0.3, -31.8555)
            (0.4, -30.5675)
            (0.5,-4.0141)
            (0.6, 3.8053)
            (0.7,23.6569)
            (0.8,42.1162)
            (0.9,  69.1649)
            (1, 99.6819)
        };
        \addlegendentry{Readout}

    \end{axis}
\end{tikzpicture}
\caption{Impact of noise sources on the accuracy of the soil sensing circuit. The phase difference between \(\phi_\text{soil}\) and \(\phi_\text{free}\) is set to 0.1 and the number of sensor qubits fixed at \(n_s = n_f = 5\).}

\label{fig:accuracy_vs_epsilon}
\end{figure}
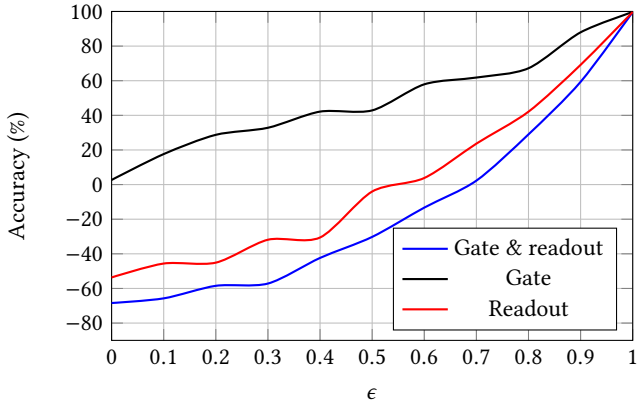

Figure \ref{fig:accuracy_vs_epsilon} examines the impact of individual noise sources by selectively enabling specific error types--readout/measurement errors (shown in red), gate errors (shown in blue, including both single- and multi-qubit gates, or both)--on both the memory and sensing qubits. To study these effects, we introduce a scaling factor, $\epsilon$, where when $\epsilon = 0$, the noise profile is set to the default, and when $\epsilon = 1$, effectively all errors for that type of noise are removed. The results indicate that gate errors have the most negligible impact on overall accuracy compared to readout errors or the combination of both, though they still contribute non-negligibly to degradation in sensing scheme performance. These findings suggest that readout errors are the dominant factor in determining the circuit's accuracy, emphasizing the importance of improving measurement fidelity in quantum sensing systems. Additional observations could delve into how these insights might inform error mitigation strategies for specific qubit platforms.

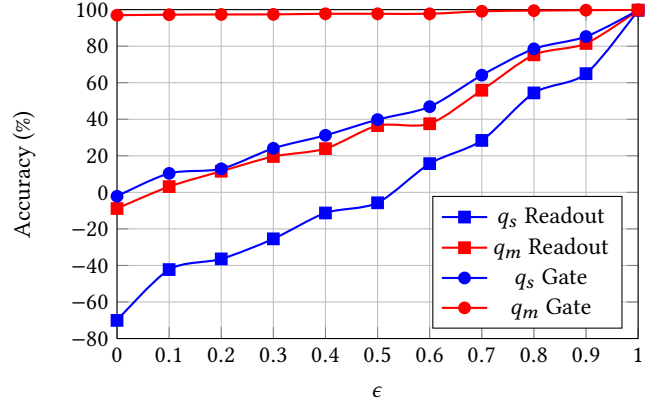
\begin{figure}
    \centering
        \begin{tikzpicture}
    \begin{axis}[
        width=\columnwidth,  
    height=0.7\columnwidth,
        xlabel={$\epsilon$},
        ylabel={Accuracy (\%)},
        xmin=0, xmax=1,
        ymin=-80, ymax=100,
        xtick={0,0.1,0.2,0.3,0.4,0.5,0.6,0.7,0.8,0.9,1},
        ytick={-80,-60,-40,-20,0,20,40,60,80,100},
        grid=both,
        minor grid style={gray!25},
        major grid style={gray!50},
        legend pos=south east
    ]
        \addplot[smooth,blue,thick, mark = square*] coordinates {
            (0, -70.0719)
            (0.1, -42.2606)
            (0.2, -36.4270)
            (0.3,-25.4490 )
            (0.4,-11.2872)
            (0.5, -5.7145)
            (0.6,15.7176)
            (0.7,28.4143)
            (0.8,54.4154)
            (0.9, 64.9619)
            (1,99.6819)
        };
        \addlegendentry{$q_s$ Readout}

        \addplot[smooth,red,thick, mark = square*] coordinates {
            (0, -8.8478)
            (0.1, 3.1505)
            (0.2,11.5531)
            (0.3, 19.6204)
            (0.4,  23.9333)
            (0.5,  36.5074)
            (0.6, 37.5464)
            (0.7,55.8891)
            (0.8, 75.2487)
            (0.9,  81.4853)
            (1, 99.6819)
        };
        \addlegendentry{$q_m$ Readout}

        \addplot[smooth,blue,thick, mark = *] coordinates {
            (0, -2.1532)
            (0.1, 10.3836)
            (0.2, 12.8736)
            (0.3,24.0999 )
            (0.4,31.2222)
            (0.5, 39.8237)
            (0.6,46.8913)
            (0.7,64.1143)
            (0.8,78.5331)
            (0.9, 85.2546)
            (1,99.6819)
        };
        \addlegendentry{$q_s$ Gate}

        \addplot[smooth,red,thick, mark = *] coordinates {
            (0, 96.9518)
            (0.1, 97.2249)
            (0.2, 97.3486)
            (0.3,97.3943)
            (0.4,97.7240)
            (0.5,  97.7265)
            (0.6, 97.7265)
            (0.7,99.0883)
            (0.8, 99.4438)
            (0.9,  99.6819)
            (1, 99.8825)
        };
        \addlegendentry{$q_m$ Gate}

    \end{axis}
\end{tikzpicture}
\caption{Impact of errors on accuracy when errors are restricted to memory or sensing qubits. \(n_s = n_f = 5\), the phase difference between \(\phi_\text{free}\) and \(\phi_\text{soil}\) is set to 0.1}

    \label{fig:enter-label}
\end{figure}

The preceding simulational results motivate a deeper analysis of how errors affect the performance of the sensing scheme when restricted to either the memory or sensing qubits. Figure \ref{fig:enter-label} shows that readout errors on the sensing qubits (blue squares) have the most significant impact on the accuracy of the radar scheme. Interestingly, gate errors on the sensing qubits (blue circles) have an effect comparable to that of readout errors on the memory qubits (red squares). In contrast, gate errors on the memory qubit (red circles) are almost negligible. These results highlight the role of the memory qubits, which have no direct rotation gates in this scheme and are only involved via CNOT operations. A memory qubit’s role in a sensing scheme is to preserve coherence and information; therefore, its performance is dominated by coherence and readout fidelity rather than the quality of rotation gate operations.

Readout errors of sensing qubits must be carefully minimized to achieve useful accuracies in the sensing scheme, as they have the most substantial impact on performance. Additionally, while gate errors on the sensing qubits have a comparable effect to readout errors on the memory qubit, the negligible impact of gate errors on the memory qubit suggests that its performance is primarily influenced by coherence and readout fidelity rather than the quality of gate operations. Thus, noise reduction efforts should prioritize improving readout fidelity and minimizing gate errors for the sensing qubits while focusing on coherence preservation for the memory qubit.

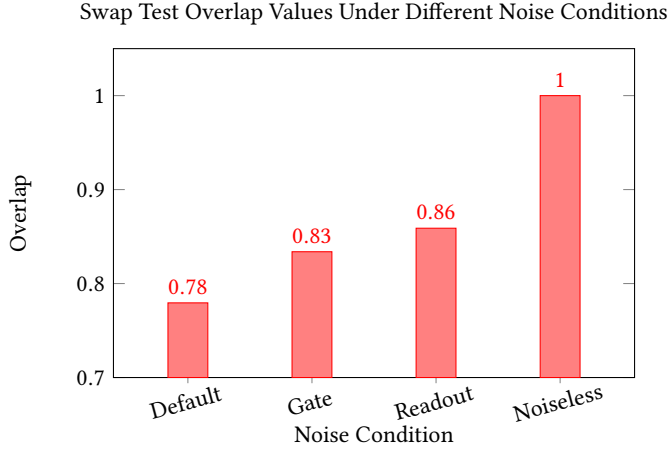
\begin{figure}[h!]
\centering
\begin{tikzpicture}
\begin{axis}[
    width=\columnwidth,  
    height=0.7\columnwidth,
    ybar,
    bar width=15pt,
    ymin=0.7, ymax=1.05,
    symbolic x coords={Default, Gate, Readout, Noiseless},
    xtick=data,
    xlabel={Noise Condition},
    ylabel={Overlap},
    nodes near coords,
    nodes near coords align={vertical},
    title={Swap Test Overlap Values Under Different Noise Conditions},
    enlarge x limits=0.2, 
    xticklabel style={
        anchor=base, 
        yshift=-10pt, 
        rotate=15 
    }, 
    xtick align=center,
    xtick pos=left 
]
\addplot[red, fill=red!50] coordinates {
    (Default, 0.779462)
    (Gate, 0.833876)
    (Readout, 0.858892)
    (Noiseless, 1.0)
};
\end{axis}
\end{tikzpicture}
\caption{Swap test results under different noise conditions for the radar application. $\phi_{\text{diff}} = 0.1$ and six total sensing probes were used for these simulations $(n_f = n_f)$ with $10^6$ shots.}
\label{fig:swap_test_results}
\end{figure}

To evaluate the performance of the similarity measure test, we employed the swap test, depicted in Figure \ref{fig:SWAP-circuit}. The simulation uses the pipeline shown in Figure \ref{fig:stqs-circuit}, with a single sensing step using ancilla computing qubits to perform the swap test. The reference state is prepared to match the state of the memory qubit under ideal, noiseless conditions. The swap test results under default noise conditions and with noise selectively removed for gate operations, readout, and in an entirely noiseless simulation are summarized in Figure \ref{fig:swap_test_results}.

\subsection{Dark Matter Detection}

Dark matter, though it makes up 85\% \cite{2020} of the universe's total mass, has remained elusive to sensing schemes despite its inferred presence through well-documented \cite{1970ApJ...159..379R, Clowe_2006} gravitational effects on galaxies and large-scale cosmological structures. A novel methodology for detecting ultra-light wavelike dark matter has recently been proposed by \cite{PhysRevLett.133.021801} that promises substantial enhancements to previous detection strategies in the 1 to 10~GHz energy range. By leveraging superconducting transmon qubits as sensitive detectors to exploit the intrinsic sensitivity of two-level quantum systems, \cite{PhysRevLett.113.220501} has theoretically demonstrated enhanced detection of dark matter signals. The proposed detection scheme can be fed into the STQS framework for analyzing, simulating, and optimizing the performance when all experimental noises, e.g., thermal relaxation, measurement, and possible depolarization errors, are considered.

Moreover, the STQS framework is particularly beneficial for dark matter detection due to its ability to establish spatial correlations, which are crucial for directional and distributed sensing schemes. As noted in \cite{Knirck_2018}, directional detection methods provide critical insights into properties such as the ``axion wind'', which is a consequence of Earth's motion relative to the dark matter halo. This directional sensitivity and the potential for enhanced spatial information further motivates our application study of dark matter detection within the STQS framework, as it provides a robust platform for exploring distributed sensing.

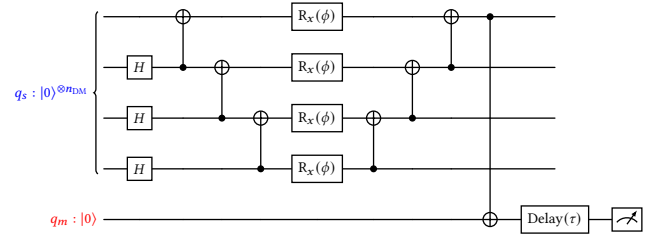
\begin{figure}
    \centering
\resizebox{\linewidth}{!}{%

\begin{quantikz}[font=\normalsize{}]
\lstick[wires=4, label style = {blue}]{$q_s: |0\rangle^{\otimes n_\text{DM}}$}& &\targ{} & & &   \gate{\text{R}_x(\phi)} & & & \targ{} &\ctrl{4} &\\
& \gate[1]{H} &  \ctrl{-1} & \targ{} & &  \gate[1]{\text{R}_x(\phi)} & & \targ{} & \ctrl{-1} & &\\
& \gate[1]{H} & & \ctrl{-1} & \targ{} & \gate[1]{\text{R}_x(\phi)}  & \targ{} & \ctrl{-1} & & &\\
& \gate{H} &  & & \ctrl{-1} & \gate{\text{R}_x(\phi)} & \ctrl{-1} & & & & \\
\lstick[label style={red}]{$q_m: {|0\rangle}$} & & & & & & & & & \targ{} & \gate[1]{\text{Delay}(\tau)} & \meter{}
\end{quantikz}

}
    \caption{Quantum circuit pipeline for one sensing step of the dark matter detection application based on proposal~\cite{PhysRevLett.133.021801}.}

    \label{fig:DM-CIRCUIT}
\end{figure}

Figure \ref{fig:DM-CIRCUIT} illustrates the quantum circuit for a single sensing step in the dark matter detection scheme, employing $n_{\text{DM}}$ sensor qubits $q_s$ and one memory qubit $q_m$. For brevity, only one sensing step is depicted; however, this pipeline can be adapted and extended to accommodate any number of sensing steps and sensor qubits by following the STQS framework outlined in Section \ref{sec:overall}. We note that extending the number of sensing steps similar to the STQS framework enables the experiments to detect the presence or absence of the wavelike dark matter and characterize potential correlations in the dark matter field. The circuit design of Figure \ref{fig:DM-CIRCUIT} mirrors the scheme initially proposed by \cite{PhysRevLett.133.021801}, adapted here for the purpose of implementing the STQS framework. Sensor qubits are prepared in GHZ-type states. GHZ-type states are employed for their well-established ability to optimize the QFI in noiseless sensing scenarios, ensuring high sensitivity to the encoded dark matter signal. However, unlike in the radar scheme, the sensed dark matter signal is encoded using $\text{R}_x$ gates rather than phase gates. After the signal is sensed, a series of disentangling gates on the sensing probes functions to balance the benefits of quantum entanglement with measurements and classical post-processing. This approach ensures that the dark matter signal is extracted with maximum fidelity while minimizing the challenges associated with directly measuring entangled states. The communication between the sensor qubits and the memory qubit is facilitated by CNOT gates, following the same mapping strategy as the radar scheme, and the memory qubit functions as a buffer, preserving the signal and enabling possible synchronization. When more than one step is needed, a delay gate is applied to the memory qubit for a duration of $\tau$, identical to its implementation in the radar scheme. This structure highlights the modularity of the STQS framework, retaining the essential features of distributed quantum sensing for ultra-light wavelike dark matter detection.

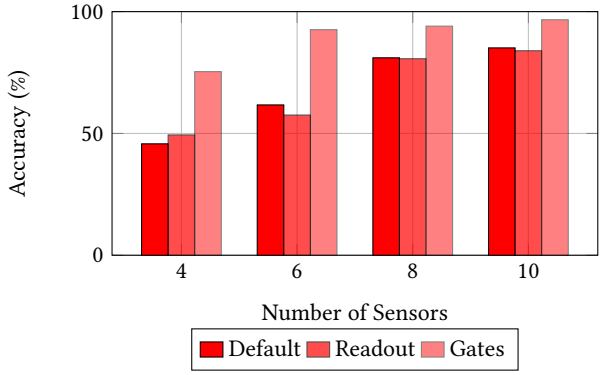
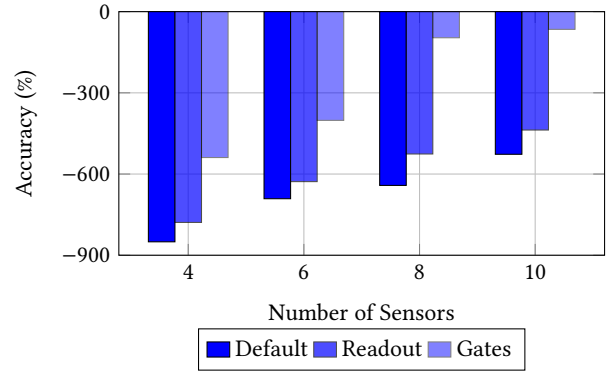
\begin{figure*}
    \centering

    \begin{subfigure}[t]{0.45\textwidth}
        \centering
        \begin{tikzpicture}
        \begin{axis}[
            width=\linewidth,
            height=0.6\linewidth,
            xlabel={Number of Sensors},
            ylabel={Accuracy (\%)},
            symbolic x coords={4, 6, 8, 10},
            xtick=data,
            ymin=0, ymax=100,
            ytick={0, 50, 100},
            bar width=10pt,
            enlarge x limits=0.2,
            legend style={at={(0.5,-0.3)}, anchor=north, legend columns=3},
            legend cell align={left},
            grid=both,
            legend image code/.code={
                \draw[#1, fill=#1] (0cm,-0.1cm) rectangle (0.3cm,0.1cm);
            }
        ]

        \addplot[
            ybar,
            bar shift=-10pt,
            fill=red
        ] coordinates {
            (4,45.73)
            (6,61.70)
            (8,81.04)
            (10,85.11)
        };
        \addlegendentry{Default}

        \addplot[
            ybar,
            bar shift=0pt,
            fill=red,
            opacity=0.7
        ] coordinates {
            (4,49.38)
            (6,57.54)
            (8,80.66)
            (10,83.93)
        };
        \addlegendentry{Readout}

        \addplot[
            ybar,
            bar shift=10pt,
            fill=red,
            opacity=0.5
        ] coordinates {
            (4, 75.36)
            (6, 92.59)
            (8, 94.04)
            (10, 96.65)
        };
        \addlegendentry{Gates}

        \end{axis}
        \end{tikzpicture}
        \caption{Data for $\phi = 0.1$.}
    \end{subfigure}
    \hfill
    \begin{subfigure}[t]{0.45\textwidth}
        \centering
        \begin{tikzpicture}
        \begin{axis}[
            width=\linewidth,
            height=0.6\linewidth,
            xlabel={Number of Sensors},
            ylabel={Accuracy (\%)},
            symbolic x coords={4, 6, 8, 10},
            xtick=data,
            ymin=-900, ymax=0,
            ytick={-900, -600, -300, 0},
            bar width=10pt,
            enlarge x limits=0.2,
            legend style={at={(0.5,-0.3)}, anchor=north, legend columns=3},
            legend cell align={left},
            grid=both,
            legend image code/.code={
                \draw[#1, fill=#1] (0cm,-0.1cm) rectangle (0.3cm,0.1cm);
            }
        ]

        \addplot[
            ybar,
            bar shift=-10pt,
            fill=blue
        ] coordinates {
            (4,-850.37)
            (6,-691.16)
            (8,-642.01)
            (10,-526.99)
        };
        \addlegendentry{Default}

        \addplot[
            ybar,
            bar shift=0pt,
            fill=blue,
            opacity=0.7
        ] coordinates {
            (4,-778.72)
            (6,-628.01)
            (8,-525.88)
            (10,-437.37)
        };
        \addlegendentry{Readout}

        \addplot[
            ybar,
            bar shift=10pt,
            fill=blue,
            opacity=0.5
        ] coordinates {
            (4,-539.07)
            (6,-401.68)
            (8,-96.54)
            (10, -65.42)
        };
        \addlegendentry{Gates}

        \end{axis}
        \end{tikzpicture}
        \caption{Data for $\phi = 0.01$.}
    \end{subfigure}

    \caption{Comparison of the sensing scheme's accuracy for different values of \(\phi\) as a function of the number of probes. Results are shown for three noise profiles, represented with varying opacity.}

    \label{fig:grouped-bar-default-noise}
\end{figure*}

Figure \ref{fig:grouped-bar-default-noise} compares different values of the sensing signal strength $\phi$ ($\phi = 0.1$ shown in red, and $\phi = 0.01$ shown in blue), examining the performance of the sensing scheme as a function of the number of probes. This analysis is of critical importance for near-term dark matter detection experiments. Since dark matter remains undetected, understanding how small the signal $\phi$ can be while still being detectable and how this can vary depending on the number of probes used provides essential guidance for experimental design. Our analysis was done following three different noise profiles (depicted with varying opacities), allowing for a thorough comparison of how noise in different parts of the sensing pipeline impacts sensing accuracy. The results illustrate that the magnitude of $\phi$ and the number of sensing probes used dramatically influence the accuracy.
Furthermore, the simulations show that for smaller values of $\phi$, there is a much greater number of sensors needed to reach the same accuracy for fewer sensors but with larger $\phi$. Interestingly, we find that turning off only readout noise while leaving gate noise active provides little to no significant improvement in accuracy over the default noise profile. This suggests that simply improving readout fidelity is insufficient to achieve high accuracy in this sensing scheme. Instead, increasing the number of probes appears to be a more effective strategy for enhancing performance. In contrast, our results indicate that eliminating gate noise leads to substantial improvement in accuracy, highlighting its appreciable contribution to the overall performance of the sensing scheme. These findings underscore the complex interplay between noise sources and sensing resources in achieving high-precision sensing accuracy. 

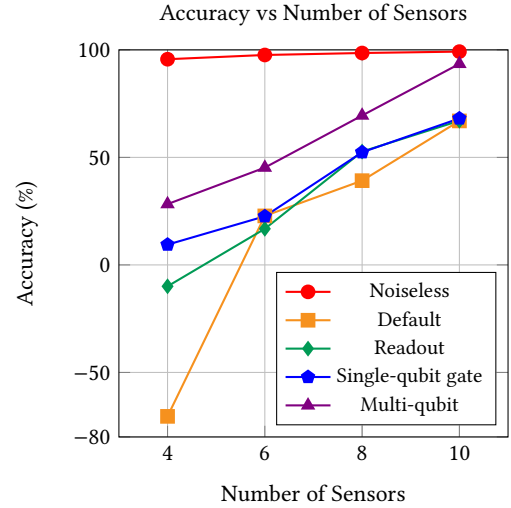
\begin{figure}
    \centering
    \resizebox{.8\columnwidth}{.8\columnwidth}{
    \begin{tikzpicture}
\begin{axis}[
    width=.8\columnwidth, height=.8\columnwidth,
    xlabel={Number of Sensors},
    ylabel={Accuracy (\%)},
    title={Accuracy vs Number of Sensors},
    xmin=3, xmax=11,
    ymin=-80, ymax=100,
    xtick={4, 6, 8, 10},
    ytick={-80, -50, 0, 50, 100},
    legend pos=south east,
    grid=both,
    legend style={font=\small}
]

\addplot[
    thick,
    red,
    mark=*,
    mark options={scale=1.2, fill=red},
] coordinates {
    (4, 95.65)
    (6, 97.62)
    (8, 98.54)
    (10, 99.19)
};
\addlegendentry{Noiseless}

\addplot[
    thick,
    BurntOrange,
    mark=square*,
    mark options={scale=1.2, fill= BurntOrange},
] coordinates {
    (4,-70.43)
    (6, 22.79)
    (8, 39.18)
    (10, 66.96)
};
\addlegendentry{Default}

\addplot[
    thick,
    ForestGreen,
    mark=diamond*,
    mark options={scale=1.2, fill=ForestGreen},
] coordinates {
    (4,-9.95)
    (6,16.79)
    (8, 52.70)
    (10,67.05)
};
\addlegendentry{Readout}

\addplot[
    thick,
    blue,
    mark=pentagon*,
    mark options={scale=1.2, fill=blue},
] coordinates {
    (4,9.43)
    (6,22.61)
    (8, 52.38)
    (10,68.12)
};
\addlegendentry{Single-qubit gate}

\addplot[
    thick,
    violet,
    mark=triangle*,
    mark options={scale=1.2, fill=violet},
] coordinates {
    (4,28.28)
    (6,45.25)
    (8, 69.47)
    (10,93.49)
};
\addlegendentry{Multi-qubit}

\end{axis}
\end{tikzpicture}
}
    \caption{Measurement accuracy for dark matter signal strength $(\phi=0.1)$ under different noise profiles as a function of the number of sensing probes.}
    \label{fig:accuracy_comparison}
\end{figure}

Figure \ref{fig:accuracy_comparison} illustrates the comparison of measurement accuracy for the dark matter signal strength $(\phi = 0.1)$, under different noise profiles, plotted as a function of the number of sensing probes used. The noiseless simulation results are shown as red circles, the default noise profile shown with burnt orange squares, readout/measurement errors completely turned off are shown as forest green diamonds, single-qubit gate errors completely turned off by blue pentagons, and multi-qubit gate errors completely turned off by violet triangles. As expected, the noiseless simulations yield the highest measurement accuracy, with a modest but present improvement as the number of sensing probes increases past four. Notably, the default noise profile exhibits the most pronounced improvement in accuracy when the number of probes increases from four to six, suggesting a greater sensitivity to probe count in this regime. For the remaining noise profiles, the scaling with respect to the number of probes appears more uniform, showing similar trends across probe counts. Interestingly, the accuracy under readout errors and single-qubit gate errors converge as the number of sensor probes reaches eight, resulting in identical performance for these two error types. This convergence shows a potential for diminishing differences between specific error mechanisms in larger sensing arrays, a feature that merits further investigation.

\begin{figure}
    \centering
    \begin{tikzpicture}
    \begin{axis}[
        width=.7\columnwidth,
        height=.7\columnwidth,
        xlabel={$\epsilon$},
        ylabel={Accuracy (\%)},
        xmin=0, xmax=1,
        ymin=-450, ymax=100,
        xtick={0,0.1,0.2,0.3,0.4,0.5,0.6,0.7,0.8,0.9,1},
        ytick={-400,-300,-200,-100,0,50,100},
        grid=both,
        minor grid style={gray!25},
        major grid style={gray!50},
        legend pos=south east,
        legend style={font=\small},
        title={Dark Matter Experiment with Noise Adjustments}
    ]

        \addplot[smooth,blue,thick,mark=*,mark options={solid,fill=blue}] coordinates {
            (0, -49.30)
            (0.1, -45.65)
            (0.2, -17.94)
            (0.3, -6.15)
            (0.4, 14.78)
            (0.5, 35.79)
            (0.6, 31.53)
            (0.7, 38.90)
            (0.8, 74.91)
            (0.9, 97.87)
            (1, 99.19)
        };
        \addlegendentry{Readout}

        \addplot[smooth,red,thick,mark=*,mark options={solid,fill=red}] coordinates {
            (0, -446.93)
            (0.1, -440.78)
            (0.2, -334.64)
            (0.3, -293.73)
            (0.4, -257.61)
            (0.5, -226.67)
            (0.6, -180.82)
            (0.7, -177.39)
            (0.8, -87.98)
            (0.9, -24.88)
            (1, 99.19)
        };
        \addlegendentry{Gates}

        \addplot[smooth,black,thick,mark=*,mark options={solid,fill=black}] coordinates {};
        \addlegendentry{Gates \& Readout}

    \end{axis}
    \end{tikzpicture}
    \caption{Accuracy for $\phi = 0.01$ with $10^6$ shots as a function of noise scaling $\epsilon$, isolating gate errors (red) and readout errors (blue).} 
    \label{fig:dark-matter-experiment}
\end{figure}
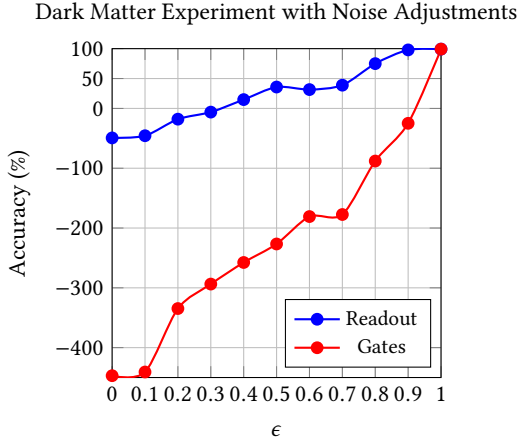

We consider a scenario with $\phi = 0.01$ and $10^6$ shots, evaluating the accuracy performance as we adjust the magnitude of specific error types. We show the results of these simulations in Figure~\ref{fig:dark-matter-experiment}. To isolate the effects of readout errors (blue) and gate errors (red), we turn off all other noise contributions and scale the remaining corresponding noise by a scale factor $\epsilon$, where $\epsilon = 0$ represents the default noise level for that type of error, and $\epsilon = 1$ corresponds to a noiseless simulation. The results show that gate errors dominate the reduction in accuracy, showing a more pronounced dampening effect than readout errors across the parameter space of all $\epsilon$. The expected trend is observed for all regimes, with accuracy improving monotonically as $\epsilon$ approaches one.

\begin{figure}[h!]
\centering
\begin{tikzpicture}
\begin{axis}[
    width=\columnwidth,  
    height=0.7\columnwidth,
    ybar,
    bar width=15pt,
    ymin=0.9, ymax=1.01,
    symbolic x coords={Default, Gate, Readout, Noiseless},
    xtick=data,
    xlabel={Noise Condition},
    ylabel={Overlap},
    nodes near coords,
    nodes near coords align={vertical},
    title={Swap Test Overlap Values Under Different Noise Conditions},
    enlarge x limits=0.2, 
    xticklabel style={
        anchor=base, 
        yshift=-10pt, 
        rotate=15 
    }, 
    xtick align=center,
    xtick pos=left 
]
\addplot[red, fill=red!50] coordinates {
    (Default,  0.920114)
    (Gate, 0.989276)
    (Readout,  0.938278)
    (Noiseless, 1.0)
};
\end{axis}
\end{tikzpicture}
\caption{Overlap values for the swap test under different noise conditions for the dark matter simulations. $\phi = 0.1$ and four total sensing probes were used for these simulations with $10^6$ shots.} 
\label{fig:swap_test_results_dm}
\end{figure}
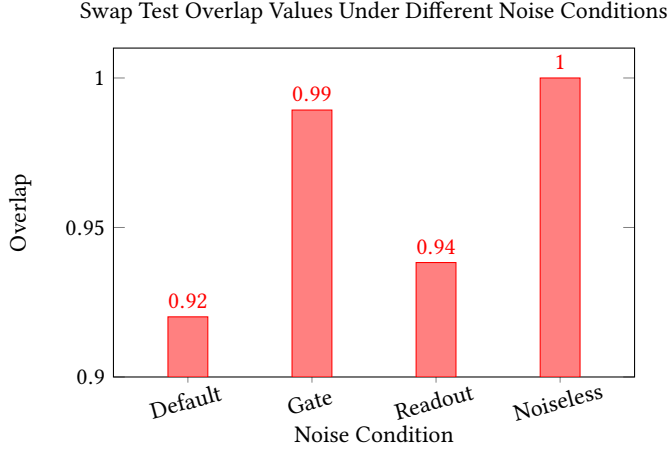

To analyze the performance of the similarity measure test,  we employed the swap test as shown in Figure~\ref{fig:SWAP-circuit}, coupled with the STQS pipeline outlined in Figure~\ref{fig:stqs-circuit}. We implemented the single sensing step as shown in Figure~\ref{fig:radar-circuit} and used ancilla computing qubits to carry out the processing. The reference state is prepared to match the state of the memory qubit in an ideal, noiseless regime. The swap test results, performed under the default noise conditions and with noise selectively turned off for gates, readout, and in an entirely noiseless simulation, are presented in Figure~\ref{fig:swap_test_results_dm}.

\section{Hardware Results}\label{sec:hardware}

We include demonstrations of quantum sensing simulations on both superconducting and trapped-ion systems, as described in Sec.~\ref{sec:ibmresult}, and Sec.~\ref{sec:ionqresult}, respectively. Both demonstrations verify our simulational findings, as detailed in Figure \ref{fig:radar-results}, validating our noise model as described in Section \ref{sec:noise model} regardless of the underlying qubit architecture. The convergence of hardware demonstrations from both systems bolsters our confidence in our noise model. It affirms its relevance for guiding future improvements in quantum sensing protocols using STQS in addition to establishing a robust foundation for possible future exploration of noise mitigation strategies in quantum sensing applications by demonstrating that the same noise mechanisms manifest across qubit hardware types.

\subsection{IBM Marrakesh}\label{sec:ibmresult}

In this section, we produce findings via demonstrations on an IBM quantum device to further substantiate our approach and analysis. Due to the limitations of the available quantum devices, we mainly focus on sensing schemes with a signal time step without using quantum memory. We use the dark matter detection scheme as our example case. We slightly modified Figure \ref{fig:DM-CIRCUIT}. To reduce resource costs and improve fidelity, we chose to omit the use of the memory qubit and instead measured the first qubit directly. These results not only confirm key aspects of our design but also illuminate opportunities for further refinement via error mitigation strategies in the context of near-term quantum devices.

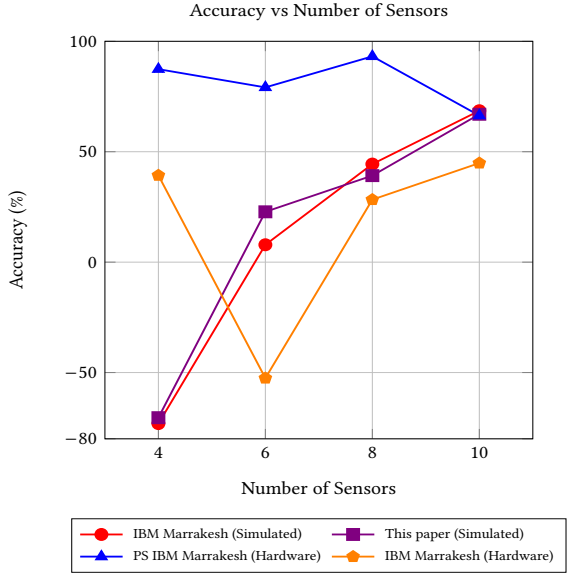
\begin{figure}
    \centering
    \resizebox{.9\columnwidth}{.9\columnwidth}{
    \begin{tikzpicture}
\begin{axis}[
    width=.9\columnwidth, 
    height=.9\columnwidth,
    xlabel={Number of Sensors},
    ylabel={Accuracy (\%)},
    title={Accuracy vs Number of Sensors},
    xmin=3, xmax=11,
    ymin=-80, ymax=100,
    xtick={4, 6, 8, 10},
    ytick={-80, -50, 0, 50, 100},
    grid=both,
    legend style={
        at={(0.5, -0.2)}, 
        anchor=north, 
        font=\scriptsize,
        legend columns = 2,
        column sep=2pt
    },
    legend cell align={left}, 
    tick label style={font=\small}, 
    label style={font=\small}, 
    title style={font=\small} 
]

\addplot[
    thick,
    red,
    mark=* ,
    mark options={scale=1.2, fill=red},
] coordinates {
    (4, -73.11)
    (6, 7.87)
    (8, 44.40)
    (10, 68.56)
};
\addlegendentry{IBM Marrakesh (Simulated)}

\addplot[
    thick,
    violet,
    mark=square*,
    mark options={scale=1.2, fill=violet},
] coordinates {
    (4, -70.43)
    (6, 22.79)
    (8, 39.18)
    (10, 66.96)
};
\addlegendentry{This paper (Simulated)}

\addplot[
    thick,
    blue,
    mark=triangle*,
    mark options={scale=1.2, fill=blue},
] coordinates {
    (4, 87.38)
    (6, 79.13)
    (8, 93.18)
    (10, 66.37)
};
\addlegendentry{PS IBM Marrakesh (Hardware)}

\addplot[
    thick,
    orange,
    mark=pentagon*,
    mark options={scale=1.2, fill=orange},
] coordinates {
    (4, 39.29)
    (6, -52.53)
    (8, 28.31)
    (10, 44.85)
};
\addlegendentry{IBM Marrakesh (Hardware)}

\end{axis}
\end{tikzpicture}
}
    \caption{Measurement accuracy for different simulated noise profiles and hardware results as a function of the number of sensors. The superconducting-based IBM device \texttt{ibm\textunderscore marrakesh} generated the hardware results. The post-selected hardware results are labeled by ``PS IBM Marrakesh''. }
    \label{fig:accuracy_comparison_real}
\end{figure}

We tested the performance of the quantum sensing circuit on IBM's \texttt{ibm\textunderscore marrakesh} device, using 10,000 shots per circuit per data set. In doing so, we highlight the practical implementation of our methods and validate their effectiveness in mitigating noise and improving accuracy in a current available quantum system. Our results are presented in Figure \ref{fig:accuracy_comparison_real}.

The four data sets shown in \ref{fig:accuracy_comparison_real}: (1) Results using the noise model presented and developed within this paper. This data set is identical to the orange-square data set in Figure \ref{fig:accuracy_comparison}. (2) Hardware-specific noise model-based simulations. This noise model is specific to the IBM device when the jobs were submitted. These simulations are in excellent agreement with the noise model presented in this paper, confirming the accuracy and applicability to actual hardware conditions. (3) Hardware results without post-selection or other error-mitigation strategies. These results are obtained directly from the measurement outcomes of the device and reflect the inherent noise and errors in the device, including decoherence and measurement errors. We postulate that the accuracy is worse than that of both simulations due to the subtle correlated noise processes within the device and environment.

The last data set in \ref{fig:accuracy_comparison_real}, (4) is hardware results with post-selection. This data set represents the hardware results after applying a post-selection technique. The post-selection is done by measuring all system qubits and only retaining measurement outcomes if all qubits without carrying sensing signal after the disentanglement stage are found in the ground state. The post-selected results significantly improve the accuracy of the measurement outcomes compared to the data set (3). Additionally, we note that this is indeed a positive outcome, as it shows that imperfect hardware can still achieve better accuracy than predicted with a low overhead mitigation technique. We suggest consideration of our results as a conservative benchmark when planning experiments and interpreting results.

\subsection{IonQ Forte}\label{sec:ionqresult}

Our IonQ Forte demonstrations seek to elucidate the performance differences between classical sensing, which employs unentangled probes, and quantum sensing, which leverages entangled probes to try and achieve enhanced measurement precision. We use 10 qubits to model the sensors in our quantum sensing demonstration. The sensing circuit is similar to Figure~\ref{fig:DM-CIRCUIT}, but without memory qubits. The sensors are prepared in an entangled GHZ-type state, enabling the coherent accumulation of quantum signals along the $X$ direction. 

For each circuit execution, we performed 2,000 measurement shots. In contrast, the classical sensing configuration uses 10 \textit{unentangled} qubits as independent probes, with each circuit execution performing 2,000 shots on each qubit. This arrangement effectively yields a statistical ensemble of 20,000 independent shots from a single sensor. Consistent with the post-selection method described in Subsection \ref{sec:ibmresult}, we only keep measurement outcomes in which all qubits except the target qubit are found in the $|0\rangle$ state to mitigate the influence of spurious measurement outcomes and improve overall fidelity. 

Limited by our hardware access, we only set $\phi=0.1$ for each qubit to simulate the signal accumulation process. We calculate the sensing accuracy to quantify the measurement results, defined as $\vert \phi_m - \phi \vert / \phi$, where $\phi_m$ is the measured angle. The classical sensing scheme results in an accuracy of 80.8\%, while the quantum sensing configuration significantly improves the accuracy to 99.1\%. This marked enhancement is not merely a consequence of the post-selection process; rather, it reflects the intrinsic advantage provided by quantum correlations. The entanglement of probes allows for the coherent accumulation of phase information, which in turn improves the sensing accuracy. 

\section{Related Work}

\subsection{Quantum Sensors}

Quantum sensors leverage quantum mechanical phenomena to achieve precision measurements, enabling breakthroughs across diverse scientific and technological domains. Platforms such as NV centers in diamond \cite{RevModPhys.96.025001, Hern_ndez_G_mez_2021, radtke2019nanoscalesensingbasednitrogen}, trapped ions \cite{RevModPhys.89.035002}, Rydberg atoms \cite{9374680, arumugam2024remotesensingsoilmoisture}, superconducting qubits \cite{danilin2021quantumsensingsuperconductingcircuits, danilin2024quantumsensingtunablesuperconducting, Fink_2024}, and photonic atoms \cite{PRXQuantum.5.010101, PhysRevX.14.031055, PhysRevResearch.3.043007} have been instrumental in advancing quantum sensing. These systems provide unique capabilities tailored to applications like magnetometry, spectroscopy, and distributed sensing, illustrating the versatility and scalability of quantum sensors. Foundational works on quantum sensing that discuss leveraging entanglement and coherence to surpass classical sensitivity limits include quantum metrological works like \cite{Giovannetti_2006}.

\subsection{Quantum Sensing Applications}

Quantum sensing has been applied to a range of impactful domain applications. These include gravitational wave detection, magnetic field sensing, biological imaging, navigation systems, and dark matter detection. Quantum magnetic and electric field sensing, has been show to enable nanoscale measurements in biomedical imaging and material characterization \cite{RevModPhys.89.015001, Rondin_2014, arumugam2024remotesensingsoilmoisture}. Gravitational wave detection enables the measurement of spacetime distortions caused by astrophysical events \cite{Abbott_2016}. Quantum radar enables the detection of low-reflectivity targets with enhanced sensitivity and is particularly useful in military and navigational applications \cite{PhysRevLett.114.080503}. Dark matter detection is a revolutionary technique of utilizing quantum systems for probing wave-like dark matter interactions \cite{PhysRevLett.133.021801, Knirck_2018}. Finally, atomic clocks, which revolutionized timekeeping with unprecedented precision, greatly impacted navigation and communication technologies \cite{Bloom_2014, RevModPhys.87.637}.

\section{Discussion}

Our work introduced a gate-based DV quantum sensing framework with customizable QSC and QSPU components to enhance signal detection for remote and distributed quantum sensing applications. We demonstrated the feasibility and scalability of this approach through simulated results of quantum radar for remote measurements of soil saturation and dark matter detection.

Our simulations underscore the significant impact of noise profiles and the number of sensor probes on accuracy scaling. Our findings indicate that increasing the number of sensor probes consistently enhances measurement accuracy, often outweighing the benefits of improving individual qubit coherence times or mitigating specific noise types. Notably, our results emphasize the pronounced advantages of scaling the sensor network size to achieve higher accuracy.

Our quantum radar and dark matter detection simulations highlight the impact of noise profiles, coherence times, and probe scaling on accuracy. For quantum radar, we found that increasing the number of sensor probes consistently improves measurement accuracy, often surpassing benefits gained from improving individual qubit coherence times or mitigating specific noise types. Similarly, probe scaling enhances measurement precision for dark matter detection, particularly for small signal strengths, where sensitivity is critical. Our results underscore the importance and advantages of distributed quantum sensing systems that prioritize probe number scaling.

Our work also revealed that readout errors on sensing qubits are the most significant limiting factor for accuracy, where these errors on memory qubits had a comparatively minor effect. Eliminating gate noise significantly enhanced performance, showing the necessity of comprehensive error mitigation strategies. Additionally, as the number of probes increased, the impact of readout and single-qubit gate errors converged, suggesting that probe scaling mitigates the influence of specific noise sources. These insights further validate the design choices of our work, specifically the preparation of memory qubits as coherence-preserving buffers.

Our application of STQS to quantum radar and dark matter detection has shown the adaptability of our framework to diverse quantum sensing protocols. The modular approach of STQS ensures compatibility with evolving sensing resources and noise mitigation technologies to help facilitate a pathway to the iterative development of quantum-enhanced detection schemes. The potential of distributed quantum sensing may open new avenues for addressing foundational challenges in physics and cosmology with significantly improved precision.

\section{Conclusion}

We have presented STQS, a unified architecture for spatiotemporal quantum sensing that integrates quantum sensing, memory, communication, and computation. By focusing on the design space and noise impacts in a distributed quantum sensing protocol, we have ensured that STQS provides practical tools for state preparation, multi-user sensing, and system scalability.

By offering an automated, scalable, and comprehensive framework, STQS contributes to tackling the long-standing challenge in distributed quantum sensing of efficiently and accurately optimizing complex experimental setups to mitigate the intrinsic noise and errors present in an experiment. Furthermore, STQS enables researchers to bypass the costly and labor-intensive processes of manually (and typically iteratively) adjusting a quantum sensing protocol's parameters so that experimental formulation is sped up and democratized by reducing the computational and resource barriers required to do quantum sensing. We believe STQS will help metrologists explore distributed quantum sensing architectures and applications that are hard to realize with current techniques. Additionally, STQS ensures that diverse infrastructures can be modeled regardless of whether the sensing scheme is for discrete or continuous-variable setups. Through our evaluations of quantum radar and dark matter detection, we have shown the capabilities of STQS-based studies to determine a better way to advance future quantum sensing scheme designs for optimal measurement accuracy.

To our knowledge, STQS is the first system-level framework for quantum sensing. As the field of quantum sensing continues to evolve dynamically, STQS will serve as a useful modeling and simulation tool to assist prototyping distributed spatial-temporal entanglement-enhance quantum sensing networks. The STQS framework can also be helpful for precision applications in fundamental physics and cosmology, given distributed quantum sensing's potential to tackle intricate detection problems with a previously unattainable accuracy level.

\section{Acknowledgments}
This research used resources of the Oak Ridge Leadership Computing Facility (OLCF), which is a DOE Office of Science User Facility supported under Contract DE-AC05-00OR22725. This research used resources of the National Energy Research Scientific Computing Center (NERSC), a U.S. Department of Energy Office of Science User Facility located at Lawrence Berkeley National Laboratory, operated under Contract No. DE-AC02-05CH11231. The Pacific Northwest National Laboratory is operated by Battelle for the U.S. Department of Energy under Contract DE-AC05-76RL01830. This simulation framework was supported by the U.S. Department of Energy, Office of Science, National Quantum Information Science Research Centers, Quantum Science Center.

\nocite{}
\bibliographystyle{ACM-Reference-Format}
\bibliography{main}

\appendix

\end{document}